\documentclass[10pt,journal,compsoc]{IEEEtran}
\ifCLASSOPTIONcompsoc
  \usepackage[nocompress]{cite}
\else
  \usepackage{cite}
\fi
\ifCLASSINFOpdf
\else
\fi
\usepackage{xspace}
\usepackage{booktabs}
\usepackage{graphicx}

\usepackage[table]{xcolor}
\usepackage{caption}
\usepackage{subcaption}
\usepackage{bbm}
\usepackage{amsmath}
\usepackage{amsfonts}
\usepackage{cleveref}
\usepackage{enumitem}

\usepackage{pifont}
\newcommand{\cmark}{\ding{51}}
\newcommand{\xmark}{\ding{55}}

\usepackage[linesnumbered,ruled]{algorithm2e}

\usepackage{rotating}
\usepackage{url}

\usepackage[normalem]{ulem}

\newcommand{\ads}{\hbox{\textsc{ADS}}\xspace}
\newcommand{\adss}{\hbox{\textsc{ADSs}}\xspace}
\newcommand{\adas}{\hbox{\textsc{ADAS}}\xspace}
\newcommand{\adass}{\hbox{\textsc{ADASs}}\xspace}

\newcommand{\apollo}{\hbox{\textsc{Apollo}}\xspace}
\newcommand{\op}{\hbox{\textsc{OpenPilot}}\xspace}

\newcommand{\carla}{\hbox{\textsc{CARLA}}\xspace}
\newcommand{\svl}{\hbox{\textsc{SVL}}\xspace}
\newcommand{\prescan}{\hbox{\textsc{PreScan}}\xspace}
\newcommand{\vtd}{\hbox{\textsc{VTD}}\xspace}
\newcommand{\paracosm}{\hbox{\textsc{Paracosm}}\xspace}
\newcommand{\airsim}{\hbox{\textsc{Airsim}}\xspace}
\newcommand{\carmaker}{\hbox{\textsc{CarMaker}}\xspace}
\newcommand{\nvidiadrivesim}{\hbox{\textsc{Nvidia Drive Sim}}\xspace}
\newcommand{\webots}{\hbox{\textsc{Webots}}\xspace}
\newcommand{\beamngtech}{\hbox{\textsc{BeamNG.tech}}\xspace}

\DeclareMathOperator*{\argmin}{arg\,min}

\begin{document}
%
\title{A Survey on Scenario-Based Testing for Automated Driving Systems \\ in High-Fidelity Simulation}
%
%
%
%

\author{Ziyuan Zhong,
        Yun Tang,
        Yuan Zhou,
        Vânia de Oliveira Neves,
        Yang Liu,
        Baishakhi Ray
\IEEEcompsocitemizethanks{\IEEEcompsocthanksitem Z. Zhong and B. Ray are with the Department
of Computer Science, Columbia University, New York,
NY, 10025.
E-mail: ziyuan.zhong@columbia.edu, rayb@cs.columbia.edu
\IEEEcompsocthanksitem Y. Tang is with Alibaba-NTU Joint Research Institute, Nanyang Technological University, Singapore. E-mail: yun005@e.ntu.edu.sg,
\IEEEcompsocthanksitem Y. Zhou and Y. Liu are with School of Computer Science and Engineering, Nanyang Technological University, Singapore E-mail: yzhou027@e.ntu.edu.sg, yangliu@ntu.edu.sg,
\IEEEcompsocthanksitem V. Neves is with Universidade Federal Fluminense, Niterói, Rio de Janeiro, Brazil. E-mail: vania@ic.uff.br
}
}

\IEEEtitleabstractindextext{%
\begin{abstract}
Automated Driving Systems (\adss) have seen rapid progress in recent years. To ensure the safety and reliability of these systems, extensive testings are being conducted before their future mass deployment. 
Testing the system on the road is the closest to real-world and desirable approach, but it is incredibly costly. Also, it is infeasible to cover rare corner cases using such real-world testing. Thus, a popular alternative is to evaluate an \ads's performance in some well-designed challenging scenarios, a.k.a. scenario-based testing. High-fidelity simulators have been widely used in this setting to maximize flexibility and convenience in testing what-if scenarios. Although many works have been proposed offering diverse frameworks/methods for testing specific systems, the comparisons and connections among these works are still missing. To bridge this gap, in this work, we provide a generic formulation of scenario-based testing in high-fidelity simulation and conduct a literature review on the existing works. We further compare them and present the open challenges as well as potential future research directions. 
\end{abstract}

}

\maketitle

\IEEEdisplaynontitleabstractindextext

%
\IEEEpeerreviewmaketitle

\IEEEraisesectionheading{\section{Introduction}\label{sec:introduction}}

Automated Driving Systems (\adss) have seen rapid progress in recent years. 
The major roadblock preventing \ads from widespread adoption is the lack of safety and functionality guarantee. To tackle this problem, many works have been conducted on safety assessment and testing of \adss. Despite the efforts that have been made, there still remain severe safety and reliability problems. Although \ads are being developed to improve safety and reduce accidents, current \ads actually introduce more accidents \cite{adstats}. It has been found that in 2020, on average, self-driving cars have 9.1 accidents per million miles driven, while the corresponding for human-driving vehicles is 4.1.

Generic testing (i.e., directly testing \ads on the road) is a widely used approach since it directly reflects the performance of the \ads under test in a real-world setting. However, from a statistical perspective, it requires more than 11 billion miles to have a 95\% confidence to ensure that an \ads is {\em only} 20\% safer than an average human driver\cite{nummilesneeded}. Besides, it is impossible to test many unknown corner cases or certain dangerous cases (e.g., a pedestrian crossing a street close to \textbf{the ego car}---the vehicle controlled by the \ads under test). As an alternative, scenario-based testing has been proposed, where testers can simulate such rare scenarios to test the ego car. It can happen either on a closed track or public road, purely in a virtual world (called software-in-loop), or a mixture of virtual and physical systems (called hardware-in-loop). Because of the flexibility and scalability, the software-in-loop approach is the most popular one, e.g., Waymo has reported more than 15 billion miles\cite{waymoreport} of virtual testing their \ads in their proprietary, high-fidelity simulator.

Over the recent years, many scenario-based testing methods in high-fidelity simulation have been proposed. Previous surveys \cite{ScenarioTestSurvey, ValidationADSafetySurvey} on \ads testing usually have a vast scope and thus only provide high-level summaries of existing works and compare them only from the algorithm's perspective.
In contrast, the current review focuses on works only on scenario-based testing in high-fidelity simulation and compares existing works by different components, including simulator, the system under test, testing objectives, scenario parameters to search, and the search algorithms. We believe the in-depth comparison provides a much clearer picture of the commonality and differences of existing works.
In summary, we make the following contributions:
\begin{itemize}[leftmargin=*,noitemsep,topsep=0pt]
    \item We study a generic framework for scenario-based testing in high-fidelity simulation.
    \item We conduct a literature review on works about scenario-based testing in high-fidelity simulation (mainly from 2018 onward) and compare them from different perspectives.
    \item We discuss the challenges and potential research directions.
\end{itemize}

The current paper is organized as the following: In \Cref{sec:background}, we introduce the background and terminology used. Next, in \Cref{sec:workflow}, we provide a generic formulation for scenario-based testing in high-fidelity simulation. We then discuss the simulators and the systems under test in \Cref{sec:system} and the scenario parameters search space in \Cref{sec:scenario}. In \Cref{sec:testing_objectives}, we introduce the testing objectives as well as the associated evaluation metrics. The selection algorithms are covered in \Cref{sec:algorithm} and \Cref{sec:algorithm_performance_estimation}. We compare the current work with the related surveys in \Cref{sec:related}. Finally, we summarize the challenges and potential directions in \Cref{sec:discussion} and conclude in \Cref{sec:conclusion}.

\section{Background}
\label{sec:background}
In this section, we introduce the background of \ads, \ads testing, and scenario, as well as our paper selection criteria. Terminologies are introduced along the way. We also provide a summary of all abbreviations used in this work in \Cref{tab:notation_table}.


\begin{table}[th]
    \centering
    \begin{tabular}{l|l}
        Notation & Meaning \\
    \toprule
    {ACC} & {Adaptive Cruise Control} \\
    ADAS & Advanced    Driver-Assistance System \\
    ADS & Automated Driving System \\
    {AEB} & {Automated Emergency Braking} \\
    {ALC} & {Automated Lane Centering} \\
    {ALC2} & {Automated Lane Change} \\
    {ARM} & {Autoregressive Model} \\
    BO & Bayesian Optimization\\
    CLS & Coverage and Local Search \\
    {CNN} & {Convolutional Neural Network} \\
    DAS & Driving Automation System\\
    DE & {Differential Evolution}\\
    DNN & Deep Neural Network \\
    DRL & Deep Reinforcement Learning \\
     DT & {Decision Tree}  \\
    EA & Evolutionary Algorithm \\
    {FCW} & {Forward Collision Warning} \\
    GA & Genetic Algorithm \\
    GP & {Gaussian Process} \\
    {IL} & {Imitation Learning} \\
    {IS} & {Importance Sampling} \\
    {LDW} & {Lane Departure Warning} \\
    {NPC} & {Non-Player Character} \\
    NN & {Neural Network} \\
    NPC & Non-Player Character\\
    ODD & Operational Design Domain \\
    {OEDR} & {Object and Event Detection and Response} \\
    {PP} & {Pedestrian Protection} \\
    {RL} & {Reinforcement Learning} \\
    SA & Simulated Annealing \\
    {TSR} & {Traffic Sign Recognition} \\
    TTC & Time to Collision\\
    
    \bottomrule
    \end{tabular}
    \caption{Table of Notations.\label{tab:notation_table}}
\end{table}

\subsection{Automated Driving Systems}
\begin{table*}[th]
    \centering
    
    \rowcolors{2}{gray!25}{white}
    \begin{tabular}{l|l|l|l|l|l|l|l}
        \rowcolor{gray!50}
        Level & Name & Example & Sustained & Lat \&  & Complete & System & Unlimited \\
        \rowcolor{gray!50}
              & & & Control & Long & OEDR & Fallback & Domain \\
    \toprule
    0 & no driving automation & automated emergency breaking & \xmark & \xmark & \xmark & \xmark & \xmark \\
    1 & driver assistance & adaptive cruise control & \cmark & \xmark & \xmark & \xmark & \xmark \\
    2 & partial driving automation & adaptive cruise control and & \cmark & \cmark & \xmark & \xmark & \xmark \\
    \rowcolor{gray!25}
    & & lane centering at the same time & & & & & \\
    \rowcolor{white}
    3 & conditional driving automation & traffic jam chauffeur & \cmark & \cmark & \cmark & \xmark & \xmark \\
    \rowcolor{gray!25}
    4 & high driving automation & local driverless taxi & \cmark & \cmark & \cmark & \cmark & \xmark \\
    \rowcolor{white}
    5 & full driving automation & driverless taxi of all conditions & \cmark & \cmark & \cmark & \cmark & \cmark \\
    \bottomrule
    \end{tabular}
    \caption{Driving automation at different levels. \label{tab:saelevels}}
\end{table*}

\begin{figure}[ht]
\centering
    \includegraphics[width=\linewidth]{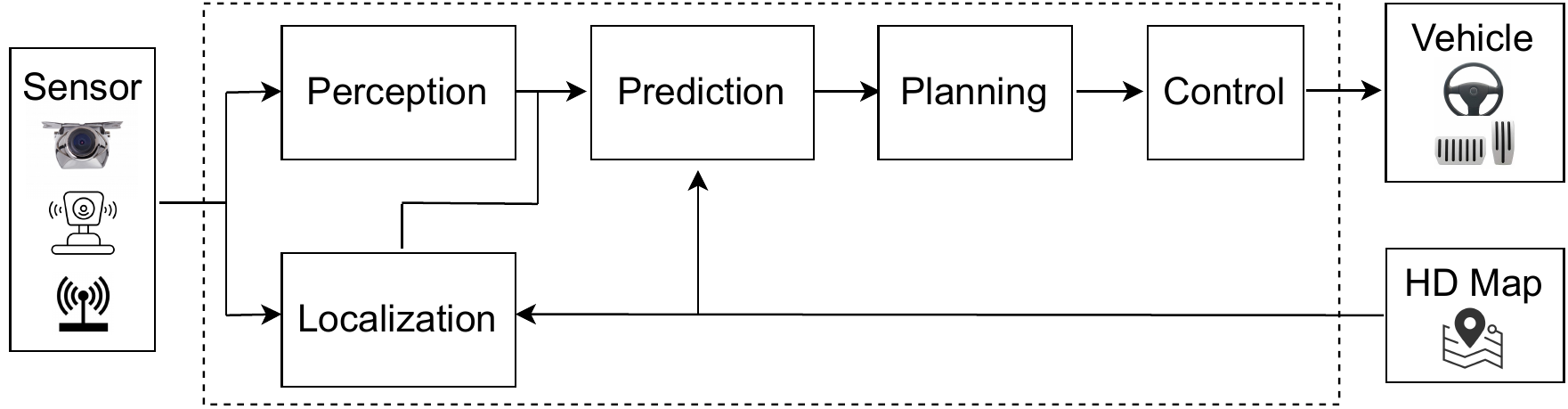}
\caption{\textbf{\small{Overview of a typical L4 \ads.\label{fig:ad_overview}}}}
\end{figure} 

As  shown in \Cref{tab:saelevels}, the
SAE J3016(TM) ”Standard Road Motor Vehicle Driving Automation System Classification and Definition”\cite{saeadlevels} 
categorizes driving automation systems into six levels from level0 to level5. We refer them as L0 to L5.
\textbf{L0} systems only perform momentary intervention during potentially hazardous situations. Typical examples include Automated Emergency Braking (AEB) and Forward Collision Warning (FCW).
\textbf{L1} systems support sustained longitudinal or latitudinal control. Typical examples include Adaptive Cruise Control (ACC), Automated Lane Centering (ALC), and Automated Lane Change (ALC2).
\textbf{L2} systems support both sustained longitudinal and sustained latitudinal driving controls at the same time. An example is a system having both ACC and ALC.
\textbf{L3} systems have the functionality of Object and Event Detection and Response (OEDR), i.e., they can monitor the driving environment and execute appropriate responses. A typical example is a traffic jam chauffeur.
\textbf{L4} systems further support system fallback, i.e.,  the system never needs human interventions within a geo-fenced region. A typical example is a local driver-less taxi. 
\textbf{L5} systems can drive under all conditions. One example is a driverless taxi that can handle any conditions. \textbf{Automated Driving System (ADS)} refers to a highly automated (i.e., L3-L5) system \cite{saeadlevels}.
\textbf{Advanced Driver Assistance System (ADAS)} describes a broad range of features, including L0 features as well as L1 features \cite{saeadlevels}. In the literature, it is commonly used to refer to an L0-L2 system.


While L5 systems, may be too ideal, many industrial companies are actively developing L4 systems. \Cref{fig:ad_overview} provides an architecture overview of a typical L4 \ads \cite{apollo, autoware}. An L4 system consumes real-time sensor data (e.g., cameras, radar, LiDAR, and GPS etc.) and HD map data to perceive its driving context including its location (\textbf{Localization}), traffic signs (\textbf{Perception}),  obstacles (\textbf{Perception}) and their trajectories (\textbf{Prediction}). The perceived driving environment along with destination information are then taken by the \textbf{Planning} module to produce short-term trajectories. Finally, the \textbf{Control} module translates the trajectory into CAN bus
\footnote{A controller area network (CAN bus) is a robust vehicle bus standard designed to allow microcontrollers and devices to communicate with each other's applications\cite{canbus}.} 
commands to control the vehicle actuators such as steering wheel and gas/brake. 

L0-L2 systems have been widely equipped in middle/high-end cars, e.g., Tesla's AutoPilot is an L2 system. L0-L2 systems usually use cameras and radar as their primary sensors and some components in \Cref{fig:ad_overview} might not be present. For example, an end-to-end lane follower (L2) might replace the entire middle block with a deep neural network (DNN). 

Because of supporting more advanced functionalities, \adss usually have very different architectures from \adass. However, we have observed that the works on testing \ads and \adas share many common features, especially when they treat the system under test as a black-box. We thus decide to include both. The works covered have tested one of L0, L1, L2, and L4 systems.


\subsection{Automated Driving Systems Testing}
\begin{figure}[ht]
\centering
    \includegraphics[width=0.8\linewidth]{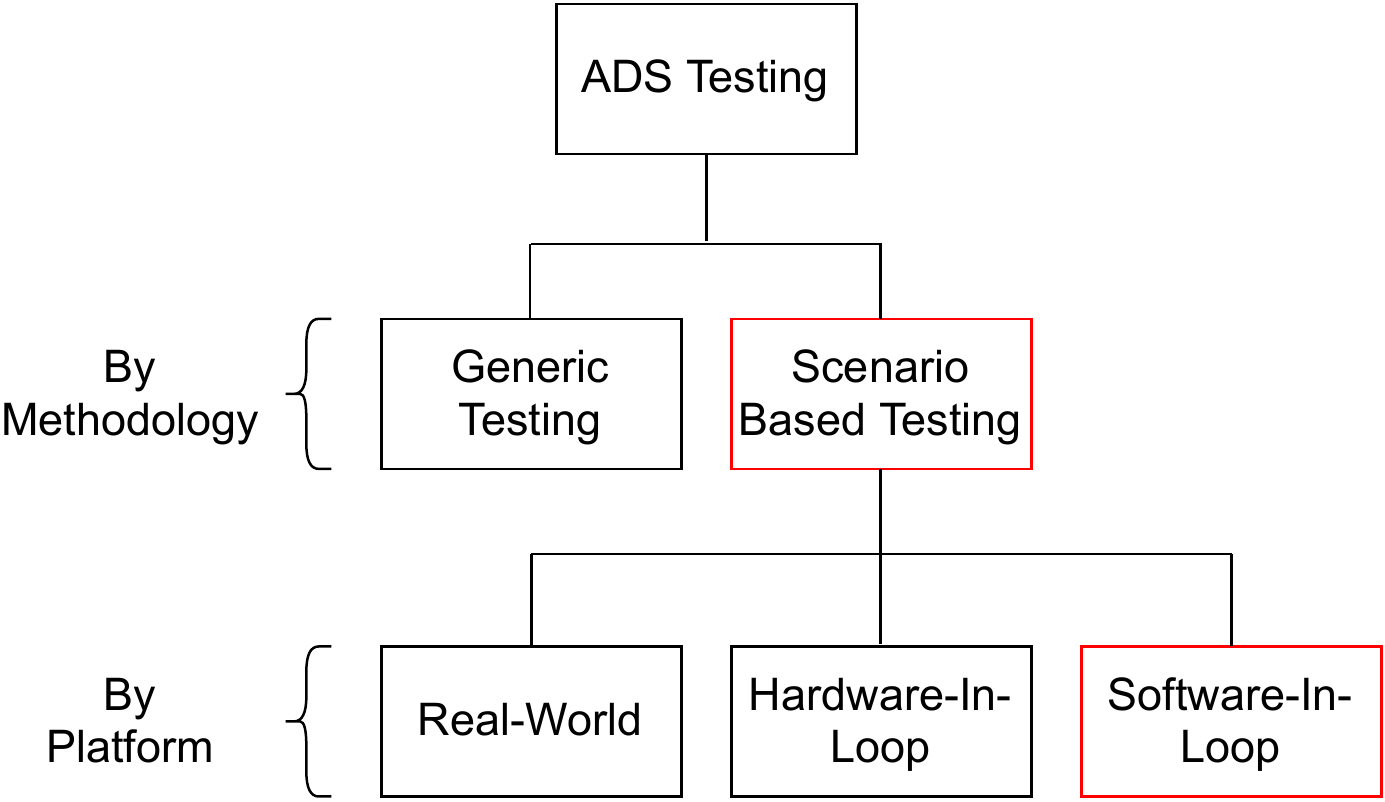}
\caption{Overview of approaches for \ads safety assessment.\label{fig:adtest_overview}}
\end{figure} 

Numerous works have been done for \ads safety assessments and the current work focuses on system-level scenario-based testing in high-fidelity simulators. In the following, we first briefly discuss the difference between component-level and system-level safety assessments, and other types of safety assessments including, formal verification and attack. Next, we introduce scenario-based testing as well as high-fidelity simulation.

\subsubsection{Component VS System Level Safety Assessment}
\textbf{Component-level safety assessment} tests individual component. For example, some works focus on testing DNNs for object detection (part of the perception component) via feeding camera images with semantic preserving variations and checking if the model's predictions are consistent \cite{shuai19, zhou19}.
The major advantage is scalability. On the flip side, the interactions among components are missed. In contrast, \textbf{system-level safety assessment} tests system-level functionalities backed by multiple components. For example, some works test if a system can reach a destination without any collisions navigating through an intersection. The tests can capture the interaction of multiple components. However, due to the complexity, techniques like formal verification are intractable. The two approaches thus complement each other to some extent. The current work focuses on system-level safety assessment and the major categories are discussed next.


\subsubsection{Other Safety Assessment Methods}
\noindent\textbf{Formal Verification}
is developed to prove the safety of a system across its entire \textbf{operational design domain (ODD)} (i.e., operating conditions under which a given \ads or feature thereof is specifically designed to function \cite{saeadlevels}) with a guarantee. 
However, methods of this category usually suffer from scalability issues and thus only focus on one component of the system  or a simple scenario with certain assumptions. 
For example, a formal model called Responsibility-Sensitive Safety (RSS) is proposed to verify a planning algorithm such that no accident can be caused for which the ego car is responsible \cite{shwartz17}. 
In order to guarantee the safety of the entire \ads, however, a system-level formal verification on all the components (rather than just a single component) is needed. 
The challenge is that the verification of some components (e.g., the perception component) has been known to be hard since they usually consist of several deep learning models that are notoriously difficult to verify \cite{shiqi2018reluval, shiqi2018neurify, wang2021betacrown}.
Consequently, no system-level verification methods have been proposed for \ads. 
Besides, it is not enough to only guarantee that the ego car can avoid the accidents for which it is responsible. To achieve safety superior to human drivers, \ads should also be able to escape from avoidable accidents caused by other non-player character (\textbf{NPC}) vehicles.

\noindent\textbf{Attack}
Another approach is attack. It assumes the existence of a malicious attacker trying to fail the ego car by tampering with either the environment or the ego car's internal states directly. 
Regarding the former, the attacker creates adversarial examples or sends malicious signals to fool the ego car's sensor processing models, e.g., perturbing front camera images\cite{Cao2019adversarial, Tian2018deeptest, Zhang2018deeproad}, road signs\cite{Eykholt2018robust, Zhao2019seeing}, rendering malicious shapes on the road\cite{sato2021dirty} or billboard\cite{Zhou2018deepbillboard}, spoofing GPS signals\cite{fusionattack}, spoofing LiDAR signals\cite{Jia2020Fooling, Wang2021AdvSimGS}, or influencing both LiDAR and camera inputs\cite{tumultisensor2021}. 
Regarding the latter, an attacker can directly inject faults inside the system to fail it\cite{rana13, chen19, jha19, fu19, jha18, rao19, abu18, pennock20, garrido20}. 
One issue with this category of methods is the assumption of an attacker's presence. Most of the time, no such attacker exists in the real world.


\subsubsection{\ads Testing}
There are two main categories of testing, i.e., generic testing and scenario-based testing. In \textbf{generic testing}, the ego car drives on public roads where the traffic scenarios are uncontrollable and unpredictable. While generic testing is necessary, it is inefficient in covering rare events involving challenging scenarios \cite{nummilesneeded}. Therefore, \textbf{scenario-based testing} has been proposed to focus on testing those challenging scenarios. 
Based on the platform, scenario-based testing can be categorized into \textbf{real-world}, \textbf{hardware-in-loop}, and \textbf{software-in-loop}. Real-world means that a scenario is constructed in the real world, which features real-life fidelity but is costly, non-scalable, and dangerous in certain scenarios. Software-in-loop, on the other hand, tests the system in a simulated environment. Software-in-loop with physical hardware is referred to as hardware-in-loop \cite{craig19, shitao19}. In an academic research environment, software-in-loop is the most popular one since it is more accessible, flexible, and scalable for scenario creation and simulation. 
The simulation environment can be simulated by a high-fidelity, photo-realistic simulator, a 2D simulator, a numeric simulation, or a dataset.  We focus on works using high-fidelity simulators as the simulation environment since numeric simulation or 2D simulators can not be used to test entire systems that take in raw sensor information. Also, dataset-based simulations, in turn,  quickly become unrealistic when the ego car deviates from the original trajectory in the dataset.

Overall, in the current paper, we review works on system-level scenario-based software-in-loop testing of \ads in high-fidelity simulators.


\subsection{Scenario}
We adopt the definition of "scenario" from \cite{simon15}. A \textbf{scenario} is a temporal sequence of scenes where a \textbf{scene} "is a snapshot of the environment including the scenery and movable objects, as well as all
actors’ and observers’ self-representations, and the relationships among those entities." Menzel et al. \cite{menzel18} extend the definition with three abstraction levels: functional, logical, and concrete. A \textbf{functional scenario} is a qualitative verbal description (e.g., the ego car drives on a straight road at high speed). A \textbf{logical scenario} is a set of parameter ranges and potentially with distributions (e.g., the ego car drives at a speed between 10m/s and 20m/s). Finally, a \textbf{concrete scenario} is a set of exact parameter values (e.g., the ego car drives at 15 m/s).

\begin{figure}[ht]
\centering
\includegraphics[width=\linewidth]{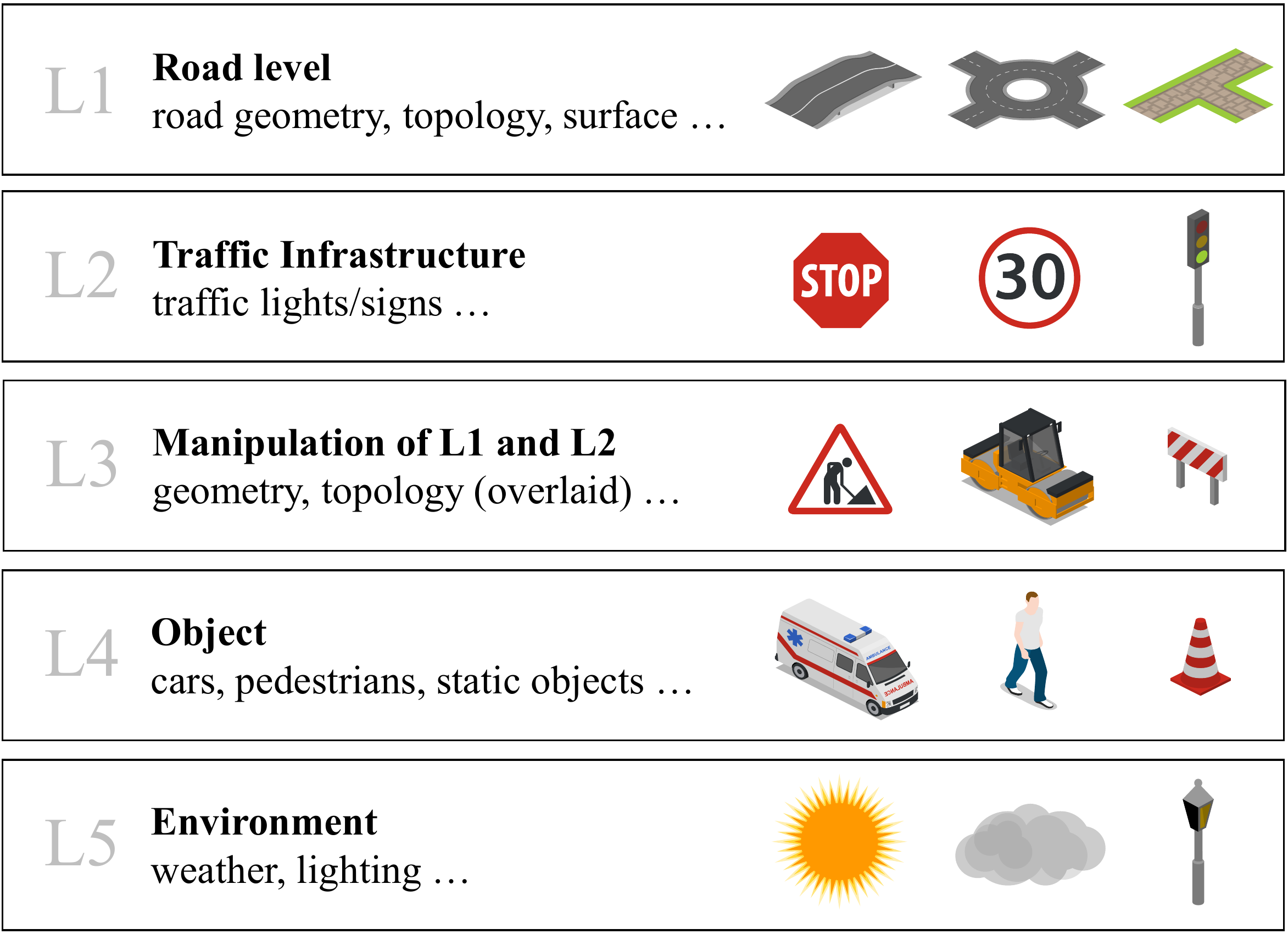}
\caption{Overview of scenario layer model.}
\label{fig:layer_model}
\end{figure}

For logical and concrete scenarios, parameters are required. Bagschik et al. \cite{bagschik18} propose a scenario layer model to structure the parameters describing a scenario into five layers according to the elements they are associated with as shown in \Cref{fig:layer_model}. In particular, \textbf{Layer1} describes the layout of the road, including markings, topology (e.g., curvature), and surface properties (e.g., friction coefficient). \textbf{Layer2} defines traffic infrastructures (e.g., traffic signs/lights). \textbf{Layer3} is about the temporary manipulation of Layer1 or Layer2. For example, a construction site affects the traffic flow. \textbf{Layer4} describes all the objects, their maneuvers, and interactions in a scenario. For example, a pedestrian crosses the street at a certain speed. Finally, \textbf{Layer5} describes the environmental condition like weather and lighting. We have classified their logical scenarios (i.e., the search space) for each work covered based on this scenario layer model.

\subsection{Paper Selection Criteria}
This section presents our paper query process, including paper selection criteria, paper sources, and query keywords.

\noindent\textbf{Inclusion Criteria:}
\begin{itemize}
    \item works on scenario-based testing for \ads and \adas in high-fidelity simulators. 
\end{itemize}
\noindent\textbf{Exclusion Criteria:}
\begin{itemize}
    \item component-level \ads safety assessment
    \item other safety assessment (e.g., formal verification, attack)
    \item works that use 2D simulators, numeric simulation, or dataset-based simulation 
    \item survey or summary papers
\end{itemize}

We collect the papers satisfying the criteria from two major sources:
\begin{itemize}
    \item major relevant conferences and journals for the last four years \footnote{We focus on the recent four years because: (1) high-fidelity simulators, especially the publicly available ones, only see rapid progress in very recent years. (2) we want to cover state-of-the-art and influential works. The state-of-the-art works are covered in source 1 and all the influential works are captured by source 2.}
    \item relevant papers citing or cited by the papers collected
\end{itemize}
The primary relevant conferences and journals include software engineering and system conferences and journals (ICSE, FSE, ASE, ISSTA, ISSRE, FASE, ICST, SOSP, OSDI, PLDI), robotics conferences and journals (ICRA, IROS, RSS), transportation conferences and journals (ITSC, IV, ICCAR, TITS, TIV), machine learning conferences and journals (Neurips, ICML, AAAI, ICLR, IJCAI), computer vision conferences and journals (CVPR, ECCV, ICCV), and security conferences and journals (CCS, USENIX, S\&P, NDSS). 

We conducted our initial search by querying using the keywords "autonomous vehicle" OR "automated vehicle" OR "autonomous driving" OR "automated driving" OR "autonomous car" OR "automated car" OR "intelligent vehicle" OR "connected vehicle" OR "self-driving" OR "self-drive" on the proceeding pages of all of the conferences and journals since these phrases are commonly used. We then went through the titles and abstracts and skimmed through the papers to determine if they satisfied the criteria. Further, we applied forward and backward
snowballing \cite{kitchenham2007guidelines, wohlin2014guidelines} on the collected papers by iteratively considering papers that are referencing or cited by the collected papers. 
For these papers, we considered them regardless of their publication year or venue. In total, we have collected 32 relevant papers shown in \Cref{tab:work_reviewed} (entries with "U" are unknown). The papers are sorted by years (shown in the column with title "Y"). For any paper that does not provide a tool name for the proposed method we use the first author's name for it. We will explain the table notations in detail when it comes to the relevant sections.




\begin{sidewaystable*}
    \centering

    \rowcolors{2}{gray!25}{white}
    \begin{tabular}{l|l|l|l|l|l|l}
        \rowcolor{gray!50}
        Y & Tool Name & Simulator & System & Testing Objective & Layer & Algorithm \\
    \toprule
16 & NSGA2-SM\cite{nsga2sm16}                 & PreScan          & FCW            & Severe (Critical,Responsible)                                & 1,4(1),5 & Sim,EA    \\
18 & FITEST\cite{fitest18}                    & PreScan          & 2*ADAS         & More (Integration-Induced),Subjective                        & 2,4(1),5 & Sim,EA    \\
18 & NSGA2-DT\cite{nsga2dt18}                 & PreScan          & AEB            & More/Severe (Critical,Responsible),Input                     & 1,4(1),5 & Sim,EA    \\
18 & Tuncali et al.\cite{tuncali18}           & Webots           & CCF            & More (Critical,Responsible),Input                            & 4(1)     & Sim,SA    \\
18 & Nitsche et al.\cite{nitsche18}           & IPG CarMaker     & 2*AEB          & Collision Rate, Task Failure Rate                             & 1,4(1)   & N,MC  \\
18 & Zhou et al.\cite{zhou18}                 & IPG CarMaker     & ACC            & Critical Boundaries                                          & 4(1)     & Sim,BO \\
19 & AC3R\cite{ac3r19}                        & BeamNG.tech      & DD             & More (Consistent with Description)                                            & U      & N,Other   \\
19 & ASFAULT\cite{asfault19}                  & BeamNG.tech      & BeamNG.AI,DD   & More (WrongLane,Diverse),Input Distance                                         & 1        & Sim,EA    \\
19 & Kluck et al.2\cite{kluck219}             & Vires VTD        & AEB            & First (Critical)                                             & 4(1)     & Sim,SA    \\
19 & Kluck et al.\cite{kluck19}               & Vires VTD        & AEB            & Severe (Critical)                                            & 4(1)     & Sim,EA    \\
19 & FAILMAKER-ADVRL\cite{failmaker-advrl19}  & AirSim           & U          & More (Critical,Natural),All                                  & 4(n)     & Step,DRL  \\
19 & Abey et al.\cite{abeysirigoonawardena19} & CARLA            & CVILLF         & First (Critical)                                             & 4(n)     & Sim,BO   \\
19 & Gangopadhyay et al.\cite{gangopadhyay19} & IPG CarMaker     & U          & More (Critical,Diverse), Input Region                        & 4(1)     & Sim,BO    \\
20 & Sim-ATAV\cite{simatav20}                 & Webots           & CMSTF          & First (Requirements-Violating)                            & 4(1)     & Sim,SA    \\
20 & AVfuzzer\cite{avfuzzer20}                & SVL              & Apollo3.5      & More (Critical, Diverse),Subjective                       & 4(k)     & Sim,EA    \\
20 & Norden et al.\cite{norden20}             & CARLA            & OPENPILOT0.5   & Critical Rate                                                & 4(n)     & Sim,IS    \\
20 & Kuutti et al.\cite{kuutti20}             & IPG CarMaker     & 2*Nav(RL, IL)  & First/More (Critical),All                                    & 4(n)     & Step,DRL  \\
20 & Ding et al.\cite{ding20}                 & CARLA            & CARLA PID      & More (Critical),All                                          & 4(2)     & Sim,ARM   \\
20 & Zhu et al.\cite{zhu20}                   & PreScan          & ACC            & More (Critical, Diverse),Input                               & 4(1),5   & Sim,Other \\
20 & Bussler et al.\cite{bussler20}           & Vires VTD        & U              & Severe (Critical)                                            & 4(2)     & Sim,EA    \\
20 & Li et al.\cite{li20}                     & Vires VTD        & AEB            & More (Critical),All                                          & 4(1)     & N,Other   \\
20 & Saquib et al.\cite{saquib20}             & CARLA            & CARLA PID      & Critical Value                                               & 4(1)     & Sim,Other   \\
21 & AutoFuzz\cite{autofuzz21}                & CARLA            & CARLA PID, LBC & More (Critical/WrongLane,Responsible,Diverse),Input Distance & 1,4(2),5 & Sim,EA    \\
21 & Paracosm\cite{paracosm21}                & Paracosm         & Nvidia CNN LF  & More (Critical, Diverse),Input Dispersion                      & 1,4(2),5 & Sim,CLS \\
21 & FusionFuzz\cite{fusionfuzz21}            & CARLA            & OPENPILOT0.8.5 & More (Critical,Fusion-Induced,Avoidable),Trajectory          & 4(k),5   & Sim,EA    \\
21 & Ding et al.2\cite{ding21}                & CARLA            & 6*Nav(RL)      & More (Critical, Diverse),All                                 & 4(2)     & Sim,Other \\
21 & Chen et al.\cite{chen21}                 & CARLA            & 2*ALC2         & More (Critical,Task Failure,Rule-Following,Diverse),All                   & 4(n)     & Step,DRL  \\
21 & ASF\cite{asf21}                          & SVL              & Apollo5.0      & More (Critical,Diverse),Subjective                           & 4(1)     & Sim,EA    \\
21 & Ghodsi et al.\cite{ghodsi21}             & Nvidia Drive Sim & U              & More (Critical),All                                & 4(2)     & N,Other   \\
21 & Tang et al.\cite{tang21}                              & SVL            & Apollo5.0      & More (Critical,Diverse),Subjective                           & 1        & N,Other   \\
21 & Tang et al.2\cite{tang221}                              & SVL            & Apollo5.0      & More (Critical,Diverse),Subjective                           & 1,4(1)   & Sim,CLS \\
21 & Tang et al.3\cite{tang321}                               & SVL            & Apollo6.0      & More (Critical,Task Failure,Diverse),Subjective            & 1,4(1)   & Sim,EA       \\                    
    \bottomrule
    \end{tabular}
    \caption{Overview of works reviewed in the current survey.\label{tab:work_reviewed}}
\end{sidewaystable*}


\section{General Workflow}
\label{sec:workflow}

\begin{figure*}[ht]
\centering
\includegraphics[width=\linewidth]{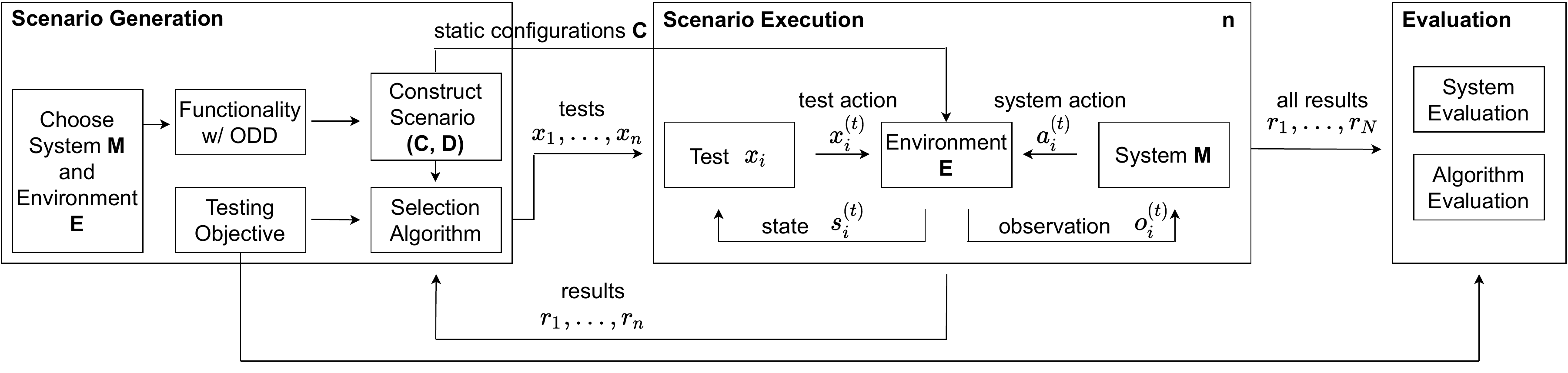}
\caption{The general workflow of scenario-based \ads testing in high-fidelity simulators.}
\label{fig:workflow}
\end{figure*}

This section summarizes a generalized workflow of all the works on scenario-based testing on \ads in high-fidelity simulators.
\Cref{fig:workflow} illustrates the overall workflow. In particular, it consists of three major components: scenario generation, scenario execution, and evaluation. We next give an overview of each component.


\subsection{Scenario Generation}
First, a system $\mathbf{M}$ under test along with the testing environment $\mathbf{E}$ needs to be chosen. An example is \op\cite{openpilot} tested in \carla\cite{carla}. Next, the functionality of the system to be tested and its ODD are specified. In this example, it can be \op's ACC functionality on a highway. A corresponding logical scenario is developed next, which consists of two parts: a static configuration set $\mathbf{C}$ and a searchable space $\mathbf{D}$. 
The static configuration $\mathbf{C}$ consists of fixed configurations and parameters, e.g., the map information and the trajectory of some background vehicles with fixed behaviors. The search space $\mathbf{D}$ consists of searchable parameters, e.g., an NPC
vehicle's speed or the parameters of a Reinforcement Learning (RL) agent which controls an NPC vehicle's behavior.

Next, a testing objective is formulated. An example is finding more scenarios causing ego car's collision given a fixed time budget. A selection algorithm that can find the parameters from the searchable space $\mathbf{D}$ to achieve the testing objective is then developed. Based on the search space and the testing objective, the algorithm can be either not adaptive, adaptive at simulation level, or adaptive at simulation step level. In other words, the algorithm can potentially update its search based on feedback at different frequencies. For any of the three categories, the selection algorithm samples a set of test parameters (a.k.a. \textbf{scenario vectors}) $d_1,...,d_n$ and uses them to parametrize test functions $x_1,...,x_n$. 
At each time step, a test function $x_i$ takes in an environment state $s_i^{(t)}$ (e.g. the locations and speed of the ego car and an NPC vehicle) and outputs a test action $x_i^{(t)}$(e.g. the acceleration for an NPC vehicle).
Finally, the test functions $x_1,...,x_n$ along with static configuration $\mathbf{C}$ are passed into the scenario execution module.


\subsection{Scenario Execution}
The scenario execution module consists of three parts: current test function $x_i$, environment $\mathbf{E}$, and system $\mathbf{M}$. All three have been provided or specified from the scenario generation module.

The test function $x_i$ first outputs the initialization test action $x_i^{(0)}$ (e.g. the initial speed of an NPC vehicle) to initialize the environment $\mathbf{E}$. At every time step $t>0$, $x_i$ takes in the current environment state $s_i^{(t)}$ and outputs test action $x_i^{(t)}$. Note that depending on the search space $\mathbf{D}$ and the algorithm, the test action at $t>0$ can either be empty or influence the environment e.g. controlling an NPC vehicle's acceleration. 

The environment $\mathbf{E}$ is initialized using the static configuration $\mathbf{C}$ as well as the initial test action provided by the test $x_i$ at the beginning, i.e., $x_i^{(0)}$. At each time step $t$, the environment $\mathbf{E}$ takes in the test action $x_i^{(t)}$ from the test function $x_i$ to potentially modify the behaviors of NPC agents, and the system action $a_i^{(t)}$ from the system $\mathbf{M}$ to control the behavior of the ego car. The environment $\mathbf{E}$ outputs the current state $s_i^{(t)}$ which usually consists of the locations and states of all objects and the observation $o_i^{t}$ which consist of the sensor information for the system $\mathbf{M}$, e.g., the front camera image.

At each time step $t>0$, the system $\mathbf{M}$ takes in the observation $o_i^{t}$ consisting of the sensor information and outputs the system action $a_i^{(t)}$ containing the control commands on the ego car in the environment.

After all the test functions finish their corresponding executions, the execution results $r_1,...,r_n$ will be returned to the algorithm that leverages this information to improve its scenario search potentially. After all the scenarios execution finishes, the results $r_1,...,r_N$ will be passed to the evaluation module.

\subsection{Evaluation}
The evaluation module consists of two parts: system evaluation and algorithm evaluation. System evaluation is about evaluating the system's performance (e.g., the collision rate of an \ads in its ODD). 
Algorithm evaluation validates the algorithm's performance, i.e., how well the algorithm achieves a testing objective. For example, it can be the number of iterations it takes for an algorithm to have a stable estimated collision rate, and the error between the collision rate and the ground-truth collision rate.

\section{Simulator \& System}
\label{sec:system}
In this section, we introduce and compare the simulators and the systems under test used in the works reviewed. Note that all the works selected leverage high-fidelity, photo-realistic simulators for their simulations, while 2D simulators like highway-env \cite{highway-env} are not within our current scope. Regarding the systems, we focus on those that are used in the papers reviewed and either are open-sourced or have enough details. In the end, we discuss the challenges and potential future directions regarding the systems and simulators.

\subsection{High-Fidelity Simulator}
Many high-fidelity simulators have been developed for \ads testing. They usually provide functionalities like 3D virtual environment creation, sensor simulation, traffic scenario creation with diverse maps and assets ready to use, and compatibility with external agents. Nearly all companies working on \ads development have their proprietary simulators for simulation-based testing to complement real-world testing. However, since they are proprietary, there is limited information regarding their details. Here we compare the simulators used in the papers reviewed from several different aspects. Note that we leave specific properties of some simulators as unknown (U) if we cannot find the details in the relevant papers and their official documentation. For details and in-depth comparison for simulators themselves, see existing surveys on simulators (e.g., \cite{simulatorsurvey1, simulatorsurvey2, simulatorsurvey3, simulatorsurvey4}).

\begin{table*}[th]
    \centering
    
    \rowcolors{2}{gray!25}{white}
    \begin{tabular}{l|l|l|l|l|l|l|l}
        \rowcolor{gray!50}
        Simulator & Source & Engine & Sensors & Assets & Map & Scenario & Compatibility \\
    \toprule
CARLA\cite{carla}            & Y    & UE4       & C L R G & V P & Urban Rural Highway & API  & Python C++ ROS1/2     \\
SVL\cite{svl}                & Y    & Unity     & C L R G & V P & Urban           & API GUI & Python ROS1/2 CyberRT \\
AirSim\cite{airsim}          & Y    & UE4       & C L G   & V P & Urban Rural & API  & Python C++ C\# Java ROS1       \\
Paracosm\cite{paracosm21}    & Y    & Unity     & C & V P & 
Urban & API  & Python                \\
Webots\cite{webots}          & Y    & ODE       & C L R G & V P & Urban Rural Highway          & GUI  & Python Matlab ROS1/2  \\
BeamNG.tech\cite{beamngtech} & N(A) & Proprietary & C L R G & V   & Urban Rural Highway       & API GUI & Python                \\
PreScan\cite{prescan}        & N    & Proprietary         & C L R G & V P & Urban Rural Highway                & API GUI  & Python C++ Matlab            \\
CarMaker\cite{carmaker}      & N    & Proprietary         & C L R G & V P & Urban Rural Highway          & API GUI  & Matlab C              \\
VTD\cite{vtd}                & N    & Proprietary         & C L R G & V P & Urban Rural Highway          & GUI  & U        \\
    \bottomrule
    \end{tabular}
    \caption{Overview of simulators covered. C:Camera, L:LiDAR, R:Radar, G:GPS, V:Vehicle, P:Pedestrian.\label{tab:simulators}}
\end{table*}

\Cref{tab:simulators} provides a comparison of all the simulators used. Note that we omitted \nvidiadrivesim\cite{nvidiadrivesim} used in \cite{ghodsi21} since it is currently in beta version and only provides minimal access and documentation. We next compare these simulators from different perspectives in the following. 

\noindent\textbf{Open Source} 
Open-source simulators are easily customized, while commercial simulators feature quality-assured technical supports. \carla, \svl (formerly known as LGSVL), \airsim, \webots, and \paracosm are open-source, and the former four have particularly active communities. Although \beamngtech (formerly known as BeamNG.research) is not open-source, one can apply free license for non-commercial use. On the other hand, \prescan, \carmaker, and \vtd require purchased licenses.

\noindent\textbf{Engine} 
The engines upon which the simulators are built directly affect the fidelity of the simulated vehicle dynamics and the rendered 3D environment. \carla and \airsim use unreal engine 4 (UE4)\cite{unrealengine}
as their physical engines. \svl and \paracosm use Unity\cite{haas2014history}. UE4 is, in general, considered more realistic visually than Unity, but there is no evidence on if it leads to a smaller gap between the simulation and the real world. Webots uses an open-source engine, Open Dynamics Engine (ODE), for rigid body dynamics simulation. The rest of the simulators are built on top of their proprietary physics engines.

\noindent\textbf{Sensors} 
The commonly implemented autonomous sensor suite includes cameras, LiDAR, radar, and GPS, supported by nearly all simulators listed. Two exceptions are \airsim, which does not have built-in support for radar, and \paracosm, which supports cameras only.

\noindent\textbf{Assets}
3D assets, i.e., vehicle and pedestrian models, are essential in constructing diverse types of scenarios.
Most simulators have a built-in vehicle and pedestrian assets ready to use except for \beamngtech, which simulates vehicles only. 

\noindent\textbf{Map}
Maps are critical components of the ODD specifications of \adss.
Most of the \adss are designed to operate in urban, rural, or highway areas, and thus, most of the simulators have constructed maps for these driving contexts. Although there are exceptions such as \svl, \airsim, and \paracosm, we believe new maps are under active construction to fill these gaps. 

\noindent\textbf{Scenario}
The usability of simulators highly depends on the interface to create and execute scenarios. Simulators like \carla, \svl, \airsim, \paracosm, \beamngtech, \prescan, and \carmaker provide API interfaces allowing users to build and execute scenarios programmably, which is essential in large-scale automated testing. In addition, some simulators, e.g., \svl, \webots, \beamngtech, \prescan, \carmaker, \vtd, also feature GUI-based scenario editors, enabling efficient scenarios prototyping.

\noindent\textbf{Compatibility}
Simulators communicate with \adss for transmitting sensor data and receiving control commands. Most of the listed simulators, e.g., \carla, \svl, \airsim, \paracosm, \webots, \beamngtech, \prescan, \carmaker provide libraries for popular programming languages such as Python, C, C++, C\#, or Java for setting up communication channels with external \adss. In addition, some simulators, i.e., \carla, \svl, \airsim, and \webots, support popular \ads communication protocols such as ROS1/2 (e.g., used by Autoware\cite{autoware}) and Apollo Cyber RT\cite{apollo} (e.g., used by Apollo), which is of great convenience for developers.

\subsection{System}
We next introduce the systems under test in the papers reviewed. Note that we only cover the publicly available systems or systems with details to reproduce. There are also other publicly available systems that can be used as the system under test, e.g., Autoware\cite{autoware}. We categorize the systems into Level 4, Level 2, and controller with ground-truth perception. 
Controllers with ground-truth perception usually take in ground-truth information of other objects and conduct path planning using methods like A*\cite{astar68} and motion control using hard-coded methods like PID\cite{pidcontrol05}.

\subsubsection{Level 4}
\noindent\textbf{\apollo} \cite{apollo} is the most popular open-source, commercial-grade L4 system developed by Baidu. It takes in sensor information from the camera, LiDAR, radar, IMU, GPS, and a high-resolution map, and has major components as in \Cref{fig:ad_overview}. It can handle many end-to-end urban driving tasks. The main reason for its popularity is that it is actively maintained and compatible with \svl. 

\noindent\textbf{Customized Multi-Sensor Trajectory Follower (CMSTF)}\cite{simatav20} consists of a perception module (taking in sensor information from the camera, LiDAR, and radar), a rule-based fusion module using unscented kalman filtering\cite{ukf00} for NPC vehicle's trajectory estimation, a planner, and a rule-based, collision avoidance controller. It has the functionality of trajectory following with collision avoidance.

\noindent\textbf{LBC}\cite{chen2019lbc} is an imitation learning based end-to-end controller. The model takes in the front camera image and a direction command (e.g., turn left) as the inputs. It follows the current lane and the direction command at the intersection. It is developed in \carla so it can be easily integrated when one chooses \carla as the simulator.


\subsubsection{Level 2/1/0}
\noindent\textbf{\op} \cite{openpilot} is an open-source commercial-grade L2 \adas developed by Comma.AI. It was ranked the first place among all commercial \adas including Tesla AutoPilot by a consumer report in 2020\cite{openpilotconsumerreport}. \op takes in the camera, radar, and IMU inputs and provides functionalities including 
ACC, ALC, FCW, and Lane Departure Warning (LDW).
Its official version provides a bridge with \carla for simple integration.

\noindent\textbf{Customized Visual Imitation Learning Trained Lane Follower (CVILLF)}\cite{abeysirigoonawardena19} is a lane follower trained using visual imitation learning. It takes in input from a front camera and provides the functionality of ACC, AEB, and ALC.

\noindent\textbf{Customized CNN Follower (CCF)}\cite{tuncali18} consists of a Convolutional Neural Network (CNN) object detector, a median flow tracker\cite{kalal10} for object tracking, and a simple collision avoidance controller. It takes in input from a front camera and provides the functionality of ACC, AEB, and ALC.

\noindent\textbf{Nvidia CNN Lane Follower (LF)}\cite{nvidiacnn} and \noindent\textbf{DeepDriving (DD)}\cite{deepdriving} are two open-source, CNN-based, end-to-end lane follower with ACC and ALC as functionalities.

Some proprietary models are denoted based on their functionalities including 
\textbf{ACC}, \textbf{AEB}, \textbf{ALC2}, \textbf{FCW}, and \textbf{ADAS} (a set of the above four).

\subsubsection{Controller with Ground-Truth Perception}


\noindent\textbf{BeamNG.AI}\cite{beamngtech} and \noindent\textbf{CARLA PID}\cite{carla} take in the simulator's ground-truth objects' information, plan a trajectory, and follow the trajectory with collision avoidance. The former is an AI agent available in \beamngtech, and the latter is a PID-based controller available in \carla.

\noindent\textbf{Nav}\cite{ding21} is a Deep Reinforcement Learning (DRL) or Imitation Learning (IL) based end-to-end controller that takes in the location and speed of all the NPC objects and outputs the control commands.

\subsection{Discussion}
\label{sec:system_discussion}

One major challenge that remains is estimating the gap between a system's testing results in a simulator and its corresponding testing results in the real world. In particular, how to quantify the gap? Which simulator and what features make the gap smaller? Do all the scenarios tend to have a similar magnitude of gap? If not, what kind of scenarios are more likely to cause inconsistent testing results? 
A natural follow-up is to mitigate the identified gap. One direction is to improve the simulator's fidelity (e.g., using a more advanced engine like Unreal Engine 5). Another direction is to make the testing objectives (see \Cref{sec:testing_objectives} for more details) have larger tolerance. For example, rather than finding only collisions in the simulation, the collision speed of the ego car at collision can be required to be larger than a threshold. The reasoning is that a severe collision found in simulation is more likely to result in a collision in the real world compare with a minor collision.

Another challenge, as can be seen from our previous discussions, is the limited availability of commercial-grade, open-source systems that are compatible with high-fidelity, open-source simulators. As a result, many proposed methods have only been shown effective on one or two sub-commercial-grade systems. Thus, a natural question is if the proposed methods in the literature can work well on those more advanced systems and how transferable their performance will be across the systems.
\section{Parameters for Scenario Search}
\label{sec:scenario}
In this section, we first discuss the parameters in the search space $\mathbf{D}$, defined in Section \ref{sec:workflow}, and used to create common scenarios in the papers reviewed. Then, we categorize the parameters according to the scenario layer model shown in \Cref{fig:layer_model}. We have found that all the works covered use a subset of parameters belonging to layer1, layer2, layer4, and layer5. We then discuss how the test functions are formulated based on the parameters selected for the search space. In the end, we discuss the observations, challenges, and potential future directions.

\subsection{Layer1}
Parameters of layer1 influence the road's layout. Typical parameters include road curvature \cite{nsga2sm16}, road ramp\cite{nsga2sm16}, road surface friction \cite{nitsche18, autofuzz21}, entire road network\cite{asfault19}, number of lanes\cite{paracosm21}, types of routes\cite{tang21}, and types of junction lanes \cite{tang221, tang321}.

\subsection{Layer2}
Parameters of layer2 influence traffic infrastructures (e.g., traffic signs). Among all the works covered, they only \cite{fitest18} allow one to control the locations of a stop/speed limit sign.

\subsection{Layer4}
Parameters of layer4 influence the objects and their behaviors. Notably, we further categorize the parameters of Layer4 into three sub-categories: \textbf{control at initialization} (denoted as \textbf{4(1)}), \textbf{control for activated behavior} (denoted as \textbf{4(2)} for one-time activated behavior parameters or \textbf{4(k)} for multiple-times activated behavior parameters), and \textbf{control at every step} (denoted as \textbf{4(n)}). Parameters belonging to control at initialization usually only influence an object's behavior at the initialization time. Parameters belonging to control for activated behavior usually define some activation criteria of an object's behavior change and its behavior after being activated. Parameters belonging to control at every step usually control an object's behavior at every time step during the simulation.

\subsubsection{Control At Initialization}
Parameters of control at initialization only influence the initial generation and behavior of an object. Typical parameters grouped by the type of objects are: 

(i) Ego Car: initial location \cite{nsga2sm16, fitest18, tuncali18, kluck219, kluck19, simatav20, saquib20}; initial velocity \cite{nsga2sm16, fitest18, kluck219, kluck19, gangopadhyay19, simatav20, zhu20, saquib20}; initial acceleration\cite{kluck219, kluck19, gangopadhyay19}; target location\cite{simatav20}; and target velocity\cite{kluck219, kluck19}.

(ii) NPC vehicle(s): number of vehicles\cite{li20}; initial location\cite{fitest18, tuncali18, nitsche18, zhou18, kluck219, kluck19, simatav20, norden20, zhu20, li20, saquib20, autofuzz21, paracosm21, ding21, asf21, tang221, tang321}; initial velocity\cite{fitest18, nitsche18, zhou18, kluck219, kluck19, simatav20, norden20, kuutti20, zhu20, li20, saquib20, autofuzz21, tang221, tang321}; initial acceleration\cite{kluck219, kluck19}; target location\cite{simatav20, autofuzz21}; target velocity\cite{kluck219, kluck19, simatav20}; max speed\cite{paracosm21}; type\cite{tuncali18, simatav20, li20, autofuzz21, fusionfuzz21}; color\cite{tuncali18, simatav20, autofuzz21, paracosm21}; lane change duration\cite{zhou18}; relative distance at lane change\cite{saquib20};
lane change distance\cite{saquib20}; and if following traffic rules and avoiding a collision\cite{autofuzz21}.

(iii) NPC pedestrian(s): number of pedestrians\cite{li20}; initial location\cite{nsga2sm16, fitest18, ding20, li20, autofuzz21}; initial velocity\cite{nsga2sm16, fitest18, gangopadhyay19, li20}; type\cite{li20, autofuzz21}; and cloth color\cite{tuncali18}.

(iv) Stationary object(s): location\cite{autofuzz21} and type\cite{autofuzz21}.

\subsubsection{Control For Activated Behavior}
Parameters of control for activated behavior control influence an object's activation condition and behavior after being activated. Typical parameters grouped by the type of objects are:

(i) NPC vehicles: velocity after certain simulation time\cite{ghodsi21}; trigger distance for activation\cite{bussler20, ding21}; trigger distance for behavior change\cite{autofuzz21}; velocity after triggered\cite{bussler20, autofuzz21}; velocity\cite{avfuzzer20, fusionfuzz21}; and lane change\cite{avfuzzer20, fusionfuzz21} at each time interval.

(ii) NPC pedestrian: trigger distance for activation\cite{ding20, bussler20, autofuzz21, paracosm21}; velocity after triggered\cite{bussler20, autofuzz21, paracosm21}.

\subsubsection{Control At Every Step}
Parameters of control at every step influence an object's behavior at every time step. Typical parameters grouped by the type of objects are:

(i) NPC vehicles: parameters for a control policy (which, when given the current state information, outputs next location \cite{failmaker-advrl19, norden20, abeysirigoonawardena19} and speed\cite{norden20, abeysirigoonawardena19}, or acceleration \cite{chen21, kuutti20})

(ii) NPC pedestrians: parameters for a control policy \cite{abeysirigoonawardena19} (which, when given the current state information, outputs next location and speed).

\subsection{Layer5}
Parameters of layer5 influence the environmental conditions like weather and lighting. These effects usually rely on a simulator's built-in weather and lighting options. Typical parameters include rain\cite{nsga2sm16, zhu20, autofuzz21, fusionfuzz21}; fog\cite{nsga2sm16, fitest18, autofuzz21, paracosm21, fusionfuzz21}; cloud\cite{autofuzz21, fusionfuzz21}; wind intensity\cite{autofuzz21, fusionfuzz21}; wetness\cite{autofuzz21, fusionfuzz21}; snow\cite{nsga2sm16}; time of a day or sun angles\cite{autofuzz21, fusionfuzz21} for lighting control; and light intensity \cite{nsga2sm16}.

\subsection{Test Functions Construction}
There are two types of test functions: initial configuration-based and state-based. 

For initial configuration based, the test action $x_i^{0}$ consists of all the parameters and exerts influence on the initialization of the environment $\mathbf{E}$. All the later test actions $x_i^{t}~\forall t>0$ are empty and have zero influence on the environment $\mathbf{E}$. Note that $x_i^{0}$ can also include parameters controlling activated behavior, which only show influence in the middle of a simulation.

For state based, $x_i^{t}$ is non-empty for all time steps $t$. For example, $x_i$ is a policy that takes in the environment state $s_i^{(t)}$ and outputs $x_i^{t}$ containing the control commands (throttle and steering angle) for a NPC vehicle.

When search space consists of control at every step of layer4, the test function is considered state-based. Otherwise, the test function is configuration-based. In the former case, the algorithm is usually RL-based, and the test function can be treated as an RL agent that decides its action based on the state. In the latter case, even if there are parameters related to activation or behavior like speed change after some simulation steps, they can be pre-defined properly in the simulator environment without changing dynamically.

\subsection{Discussion}
\label{sec:scenario_discussion}
One observation is that very few works so far have made efforts to test layer2, and no work has ever tested layer3. However, parameters associated with these layers are very likely to cause a system to malfunction. For example, the placement of a temporary construction site in the middle of a highway (layer3) might cause an \ads's incorrect localization results due to the conflicts between its high-resolution map and the LiDAR sensor input. 

Another observation is that most works play with the parameters of Layer4 to control other NPC vehicles' movements during a simulation. However, there is no consensus regarding how to parametrize them. In terms of a spectrum of granularity, one end is the exact movement control of a vehicle at every time step, and the other end is only setting up its initial location and speed. They trade-off between the dimensionality of the search space and the granularity of objects' movements. It might be worth exploring what parameterization is at the "sweet spot" regarding the trade-off such that a given testing objective is best achieved.
\section{Testing Objectives \& Evaluation Metric}
\label{sec:testing_objectives}
In this section, we discuss and compare the testing objectives as well as their associated evaluation metrics (for both system and algorithm). Since all of the testing objectives involve finding a list of scenarios possessing a set of properties (\Cref{sec:property}), we first categorize and analyze these properties. Next, we introduce the two major categories of testing objectives: \textbf{special scenarios search} (\Cref{sec:special_scenarios_search}) and \textbf{performance evaluation} (\Cref{sec:performance_estimation}). Regarding the former, the goal is to find scenarios satisfying specific properties efficiently. It can be further categorized into finding more qualified scenarios (denoted as \textbf{more}), finding scenarios with the best values (denoted as \textbf{severe}), and quickly finding the first qualified scenario (denoted as \textbf{first}). Regarding the latter, the goal is to estimate the ego car's certain performance measure in its ODD efficiently. It can be further categorized into key-value estimation and critical boundary estimation.

The values for each entry in the corresponding column in \Cref{tab:work_reviewed} are as the following. When the category is special scenarios search, it is \textbf{"severe/more/first,(properties of the scenario to find),[criteria for counting when "more" is used]"}. When the category is performance estimation, it is one of \textbf{"collision rate"}, \textbf{"task failure rate"}, \textbf{"critical rate"}, \textbf{"critical value"}, and \textbf{"critical boundary"}, corresponding to the value/function to estimate.


\subsection{Properties Overview}
\label{sec:property}

\begin{figure}[ht]
\centering
    \includegraphics[width=\linewidth]{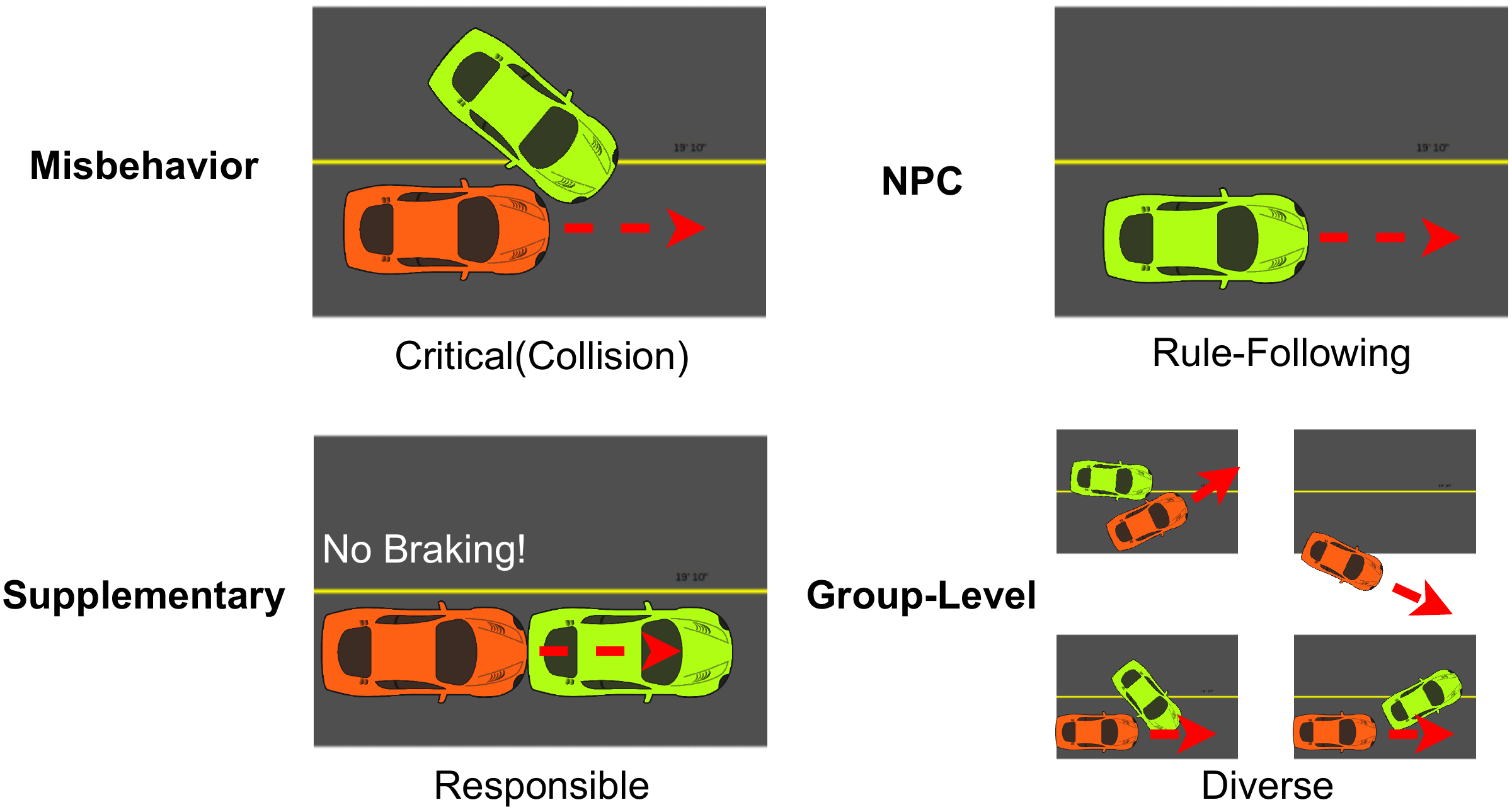}
\caption{Representative properties of each category. The orange car is the ego car and the green car is an NPC car. \label{fig:properties}}
\end{figure} 

\Cref{fig:properties} shows some representative properties of each category of properties. We next provide the details of all properties.

\subsubsection{Misbehavior Property}
These properties represent the ego car's certain misbehaviors.

\noindent\textbf{Critical} is the most well-studied property in literature. A scenario is considered critical if a safety measure is below a user-specified threshold. Two of the most commonly used measures are the minimum distance between the ego car and any other objects\cite{nsga2sm16, tuncali18, nitsche18, zhou18, failmaker-advrl19, abeysirigoonawardena19, gangopadhyay19, avfuzzer20, kuutti20, ding20, bussler20, autofuzz21, paracosm21, fusionfuzz21, ding21, chen21, asf21, tang21, tang221, tang321} during the simulation (denoted as \textbf{min distance}) and minimum Time To Collision\cite{ttc13} (or its variants) between the ego car and any of other objects \cite{nsga2sm16, nsga2dt18, kluck219, kluck19, zhu20, bussler20, li20} during the simulation (denoted as \textbf{min TTC}). Many works (e.g.\cite{zhou18}) only consider a critical scenario defined as the ego car having a collision, which is a special case of min distance with the threshold set to 0.

In practice, the min distance is usually measured using the two objects' centers or their closest surfaces. Existing works approximate the latter using the former subtracted by the objects' half lengths. This approximation is usually considered sufficient when not many lane changes happen, in which case the vehicles' widths and relative angle might also need to be considered. The min TTC measures the minimal time required for a vehicle to hit an object if both continue with the same speed and do not change their paths. 
Several more fine-grained measures are proposed in \cite{ghodsi21} which measure the effort of the ego car to escape from a potential collision by measuring the area of the ego car's potential escape routes on a discretized 2D plane. 

\cite{nsga2dt18, autofuzz21, paracosm21} use the ego car's speed during a collision to approximate the severity of a critical scenario. This differs from the previous ones by approximating the severity of a collision rather than how close the ego car is to a collision.

\noindent\textbf{Wrong Lane} scenarios are those in which the ego car drives to a wrong lane (including off-road) \cite{asfault19, autofuzz21}. The property can be quantified by the ego car's center's smallest distance to the closest valid region's boundaries. When this value is equal to zero, the ego car is considered to drive to an invalid region.

\noindent\textbf{Task Failure}
scenarios are those where the ego car fails to reach its original destination which is usually due to a deadlock with an NPC vehicle \cite{nitsche18, tang321, chen21}.

\subsubsection{Supplementary Property}
These properties usually serve as additional properties for some misbehavior properties (usually "critical").

\noindent\textbf{Avoidable} scenarios are those where if the ego car reacts properly, a critical scenario can be avoided \cite{fusionfuzz21}. 

\noindent\textbf{Responsible} scenarios are those where the ego car should take at least partial responsibility for inducing a critical scenario. In the literature, responsible properties typically considered are that the object which the ego car collides with must be in front of the ego car\cite{tuncali18}, 
within the ego car's sensors range \cite{nsga2sm16, autofuzz21}, or detected by the ego car's sensor \cite{nsga2dt18}. \cite{autofuzz21} requires the ego car to have positive speed when a collision happens. 

\noindent\textbf{Fusion-Induced} scenarios are those in which a critical scenario is caused by the malfunction of a system's fusion component, which fuses the predictions of multiple sensors (e.g., camera, radar) on some variable, e.g., the location of an NPC vehicle, and outputs a fused prediction on that variable \cite{fusionfuzz21}.

\noindent\textbf{Integration-Induced} scenarios are those in which the system's misbehaviors are induced by an integration component that decides which upstream functionality (e.g., AEB, ACC, PP, and TSR) command to use \cite{fitest18}. In particular, an integration-induced error is defined to be the one that, due to a wrong choice of the integration component, a certain safety rule (including no collision with any vehicle/pedestrian, stopping at a stop sign, respecting speed limit, and respecting safety distance) is violated.

\subsubsection{NPC Property}
These properties describe the behaviors of the NPC vehicles and usually also contribute to the responsibility of the ego car when an accident happens.

\noindent\textbf{Rule-Following}
scenarios are those in which the NPC vehicles must follow the traffic rules \cite{chen21}. 

\noindent\textbf{Natural}
scenarios are those in which all NPC vehicles do not collide with each other or walls and arrive at the destination on time \cite{failmaker-advrl19}.

\subsubsection{Group-Level Property}
Unlike other properties which can be checked independently, a group-level property depends on the relationship among a set of scenarios.

\noindent\textbf{Diverse}
is also usually used together with "critical". Many works try to find a set of diverse scenarios. However, there is no consensus regarding the definition of diversity. Diversity can be defined based on the scenario parameters, environment state, or system state during a simulation. \cite{nsga2dt18, tuncali18, ac3r19, zhu20} define diverse scenarios to be those that have different scenario parameters. \cite{asfault19, autofuzz21} define those scenario parameters to be a certain distance away in a metric space. \cite{paracosm21} defines diversity based on the overall dispersion of the scenario parameters. \cite{gangopadhyay19} defines diversity based on the distinct local minimum in the search space with respect to a critical value. In contrast, \cite{asf21, avfuzzer20, asf21, tang21, tang221, tang321} define diversity based on a subjective judgment of the environment state of the simulations. \cite{tang21, tang221, tang321} additionally consider diversity based on simulation route/junction lane coverage.

\subsubsection{Other Property}
\label{sec:other_property}
\noindent\textbf{Requirements Violating}
scenarios are those that violate a subset of a set of user-specified  requirements, which includes ego car should not collide, the sensor can detect visible objects within a specified time, localization error should not be too large for too long, the sensor-related fault should not lead to a system-level fault, and the vehicle does not brake unnecessarily or too often \cite{simatav20}.

\noindent\textbf{Consistent with Description}
scenarios are those reconstructed exactly following the descriptions in the selected police reports \cite{ac3r19}.

\subsection{Special Scenarios Search}
\label{sec:special_scenarios_search}
\subsubsection{More qualified scenarios}
Let $\phi$ denote a specific condition during the simulation (e.g., the ego car collides with an NPC vehicle). Then, the goal of this category is to maximize the number of found concrete scenarios satisfying the condition:
\[
\mathbf{X}_{\textrm{fail}} = \{\mathbf{x} \in \mathbf{X} | f(\mathbf{x}) \in \phi\}
\]

\noindent\textbf{Algorithm Evaluation Metric} is usually the number of condition satisfying scenarios found. An open question is how to count this number. The counting is usually closely related to the diversity property used. One can count the number of runs satisfying the conditions \cite{failmaker-advrl19, kuutti20, ding20, li20, ding21, ghodsi21} (denoted as \textbf{All}). Alternatively, one can base on subjective judgement \cite{fitest18, avfuzzer20, asf21, tang21, tang221, tang321} (denoted as \textbf{Subjective}); distinct spatial and speed trajectory of the ego car\cite{fusionfuzz21} (denoted as \textbf{Trajectory}); distinct scenario parameters\cite{nsga2dt18, tuncali18, ac3r19, zhu20} (denoted as \textbf{Input}); distinct scenario parameters regions\cite{gangopadhyay19} (denoted as \textbf{Input Region}); distinct scenario parameters with certain distance among them \cite{asfault19, autofuzz21} (denoted as \textbf{Input Distance}); or scenario parameters dispersion\cite{paracosm21} (denoted as \textbf{Input Dispersion}). Other than purely counting numbers, distribution of $f(x)$ for all the run cases, the average critical values or collision rates for all the runs have also been widely used.

\noindent\textbf{System Evaluation Metric} is the same as algorithm evaluation metric since it reflects, given a fixed search budget, the number of qualified scenarios a system is found to have. Since the qualified scenarios are usually those where the system misbehaves, the more the system has, the worse it performs. However, a caveat is that the search algorithm may be good at finding qualified scenarios for some systems but not for others. Thus, the number cannot be directly used to compare different systems' performance unless a certain coverage criterion has been achieved during the search process. This caveat also applies to other subcategories under special scenario search.

\subsubsection{Scenarios with the best values}
The second category is to find a concrete scenario that has a minimum value of $f(x)$ (e.g., min TTC):
\[
\mathbf{x}_{\textrm{min}} = \argmin_{\mathbf{x}} f(\mathbf{x})
\]

Note $f(x)$ can potentially be a multi-dimensional vector, in which case the goal becomes finding the best Pareto front.

\noindent\textbf{Algorithm Evaluation Metric} is usually the minimum value itself e.g. min distance/min TTC\cite{kluck19, bussler20}. In the case of multiple objectives, hypervolume\cite{nsga2sm16, nsga2dt18}, generational distance\cite{nsga2dt18}, and spread\cite{nsga2dt18} have been used.

\noindent\textbf{System Evaluation Metric} is the minimum value $f(x_{\textrm{min}})$ found which reflects the known worst-case performance of the system.

\subsubsection{Quickly finding the first qualified scenario}
The third category is to find the first qualified scenario  
\[
\mathbf{x} ~s.t.~ f(\mathbf{x})\in\phi
\]
using the smallest number of simulations.

\noindent\textbf{Algorithm Evaluation Metric} is the time/number of simulations it takes to find the first qualified scenario\cite{kluck219, abeysirigoonawardena19, simatav20, kuutti20}.

\noindent\textbf{System Evaluation Metric} is the same as the algorithm evaluation metric. Usually, the faster a qualified scenario is found, the easier it is found for the system under test.

\subsection{Performance Estimation}
\label{sec:performance_estimation}

\subsubsection{Key-Value Estimation}
\cite{norden20, nitsche18} aim to estimate the probability of a system's collision/critical/task failure rate given an ODD.
\[
\mathbf{P}_{\textrm{fail}} = \mathbb{E}[\mathbbm{1}\{f(\textbf{x})\not\in \phi \}]
\]

Similarly, \cite{saquib20} estimates the criticality value (e.g., TTC) of the ego car given an ODD.

\noindent\textbf{Algorithm Evaluation Metric} is the number of iterations to achieve a stable estimation and the error of the estimation from the ground-truth.

\noindent\textbf{System Evaluation Metric} is the estimated key-value that reflects the ego car's average performance in the ODD.

\subsubsection{Critical Boundary Estimation}
\cite{zhou18} tries to find a safety-critical boundary in the search space. In other words, it finds a function $g$ such that:
\[ 
  g(x) = 
  \begin{cases} 
  1, & \textrm{if }  f(\mathbf{x})\in \phi \\ 
  0, & \textrm{otherwise} 
  \end{cases} 
\]

The main challenge is that the computational complexity grows exponentially as the dimensions of the search space goes up.

\noindent\textbf{Algorithm Evaluation Metric} is the same as that in Key Value Estimation.

\noindent\textbf{System Evaluation Metric} is the estimated critical boundary.

\subsection{Discussion}
\label{sec:testing_objectives_discussion}
One question remains is regarding the choice of properties to be searched for. Existing works usually have their own requirements and reasoning regarding the inclusion or exclusion of certain properties. Some properties like "critical" are widely studied. Others like "avoidable" or "natural" have received much less attention. However, such properties might also be essential for the found scenarios to be more realistic and valuable. Besides, the properties might have trade-offs in practice. Exploration regarding the proper use of each property and its trade-offs can provide insights on when to use each and how to weigh each during the search process to meet one's requirements.

A closely related question is how to quantify each property such that the found scenarios are as expected. \cite{surrogate17} provides a comprehensive review on different safety measures for "critical" along with the discussions of their pros and cons. It might be worth comparing the effects of using these different measures in the context of scenario-based testing in high-fidelity simulators.

Another question is about the definition of "diversity" both as a property and as the basis for the counting standard in "finding more qualified scenarios". Any diversity criteria can be formulated based on all the state information of test function, environment, and system from each simulation. If no universally accepted definition can be defined, then under what contexts each definition should be used needs to be explored. The concept of "diversity" is also closely related to the concept of "coverage". Finding diverse scenarios (based on certain criteria) during the search process is similar to achieving certain coverage criteria. Although there is no consensus regarding a good definition of coverage, \cite{coverage20} has conducted a high-level survey on existing literature about maximizing different coverage criteria. 
One potential direction along this line is to explore the impact of using different diversity/coverage criteria and justify the use of some of them under different settings.

\section{Algorithm for Special Scenario Search}
\label{sec:algorithm}
In this section, we introduce the algorithms used to search parameters in $\textbf{D}$ for special scenario search (\Cref{sec:special_scenarios_search}). 
We first categorize the algorithms based on how adaptive they are into three categories: \textbf{non-adaptive}, \textbf{simulation-based adaptive}, and \textbf{step-based adaptive}, as shown in \Cref{tab:algorithm_categories}. 
A non-adaptive algorithm generates all the test functions at once. The algorithm does not need to access the information of the environment or system at all.
A simulation-based adaptive algorithm generates a batch of test functions at a time and passes them to the scenario execution module to run. It then leverages the feedback to update its current state off-line and generates the next batch.
Finally, a step-based adaptive algorithm usually generates one test function and constantly takes in feedback from the environment to update the current state of the test function at each simulation step.

\begin{table}[th]
    \centering
    \begin{tabular}{l|l|l|l}
        \rowcolor{gray!50}
        Category & Update & Access & Algorithms\\
    \toprule
    Non & No & No & \cite{nitsche18, ac3r19, li20, ghodsi21, tang21}\\
    \midrule
    Simulation Based & Off-line & Low & \cite{nsga2sm16, fitest18, nsga2dt18, tuncali18, zhou18} \\
    & & & \cite{asfault19, kluck219, kluck19, abeysirigoonawardena19, gangopadhyay19} \\
    & & & \cite{simatav20, avfuzzer20, norden20, ding20, zhu20} \\
    & & & \cite{bussler20, saquib20, autofuzz21, paracosm21, fusionfuzz21} \\
    & & & \cite{ding21, asf21, tang221, tang321} \\
    \midrule
    Step Based & Online & High & \cite{failmaker-advrl19, kuutti20, chen21} \\
    \bottomrule
    \end{tabular}
    \caption{Comparison of different algorithm categories.\label{tab:algorithm_categories}}
\end{table}

\subsection{Non-Adaptive}
Non-adaptive algorithms generate all test functions at once based on certain criteria, e.g., being consistent with descriptions in police reports or maximizing routes coverage. 

\cite{ac3r19} applies standard natural language processing techniques to extract the conditions of the environment, road, and vehicles from police reports. It then plans the trajectories of involved vehicles to satisfy their extracted conditions as well as vehicle kinematics, and finally reconstructs the concrete scenarios based on the trajectories and the conditions. The underlying reasoning of using police reports is that an \ads is also likely to crash when the scenarios are dangerous for human drivers.
\cite{li20} formulates an ontology and uses it to generate critical scenarios for an AEB in a car following scenario. To reduce the total number of generated scenario vectors, the algorithm flattens the ontology before applying t-wise combinatorial testing\cite{twise04}. To avoid generating invalid scenarios (e.g., NPC vehicles with overlapping initial positions), it imposes constraints during the generation process. 
\cite{ghodsi21} tries to generate more critical scenarios by directly changing the policy of a couple of vehicles closest to the ego car for three seconds by making them rush to the ego car via setting their orientation and acceleration. 
\cite{tang21} also aims to find more critical scenarios. It proposes a new route coverage metric to categorize routes on a given map and then tests \apollo on the representative routes from each category. 

\subsubsection{Discussion}
\label{sec:algorithm_non_adaptive_discussion}
Non-adaptive algorithms generate challenging scenarios based on different sources, e.g., police reports, expert-designed ontology, and dangerous actions or route coverage based on domain knowledge. All of these sources have pros and cons. For example, police reports\cite{ac3r19} cover the common difficult scenarios for human drivers, but they are not necessarily as difficult for \adss. Expert-designed ontology\cite{li20} heavily relies on the expert's knowledge. The design also needs to balance completeness and computational feasibility. Expert-designed dangerous actions\cite{ghodsi21} can create critical scenarios easily but at the same time are limited to the situations in which other vehicles are aggressive. Route coverage\cite{tang21} focuses only on the route level but does not consider the behaviors of other objects. Because of the different pros and cons, one potential direction is to combine multiple sources to generate scenarios that address the issues from any single source.

\subsection{Simulation-Based Adaptive}
Simulation-based adaptive algorithms usually run for several rounds. At each round, they learn from the feedback of a batch of simulation runs in the scenario execution module and generate a new batch of scenarios to run. Based on the methodology used, they can be categorized into \textbf{Evolutionary Algorithm} (\textbf{EA}), \textbf{Simulated Annealing} (\textbf{SA}), \textbf{Autoregressive Model} (\textbf{ARM}),
\textbf{Bayesian Optimization} (\textbf{BO}), \textbf{Coverage and Local Search} (\textbf{CLS}), and \textbf{Other}.

\subsubsection{Evolutionary Algorithm}
\label{sec:ea}
Evolutionary algorithm is popular because it is intuitive, easy to implement, and effective even if the search space is high-dimensional and the number of data points is relatively small. On the flip side, EA lacks any theoretical convergence guarantee and sometimes requires extra efforts to tune the hyper-parameters. There are two types of EA used in the literature: Genetic Algorithm (GA) and Differential Evolution (DE). We first show the routines of a typical EA and then introduce how the literature design each step in the context of \ads testing.

\begin{figure}[ht]
\centering
\includegraphics[width=0.8\linewidth]{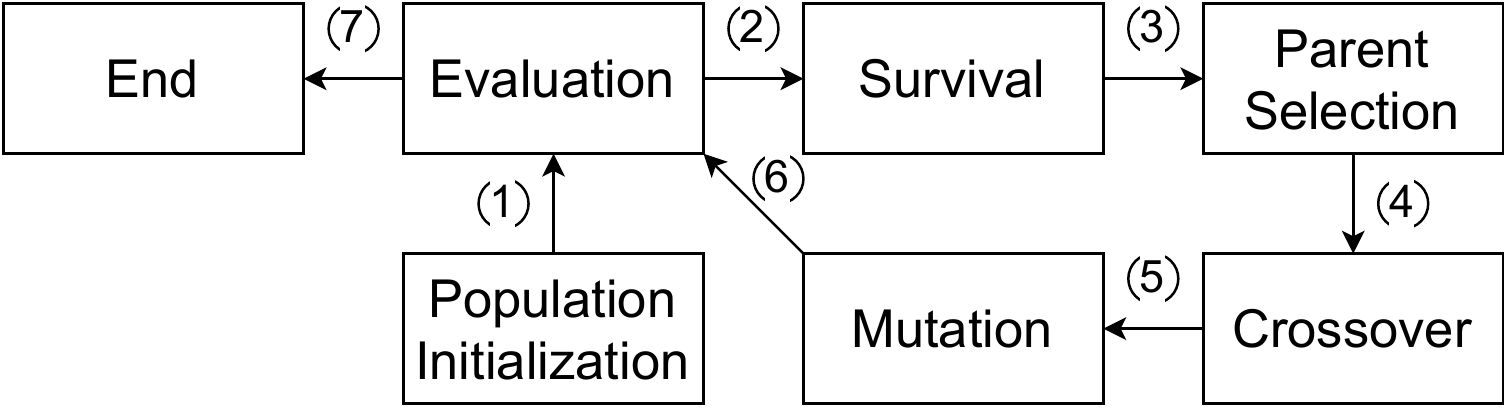}
\caption{The general workflow of a typical EA.}
\label{fig:ea}
\end{figure} 

As shown in \Cref{fig:ea}, an EA first generates initial scenario vectors and evaluates them. Each generation repeats the process of selecting parent scenario vectors, applying crossover and mutation on the chosen scenario vectors, and evaluating the fitness scores of the mutated scenario vectors. Existing works using EA differ in how they adapt each procedure for achieving their particular testing objectives. 


\noindent\textbf{Initialization}
A good initialization function for EA  can usually generate a diverse initial population. 
Most works use random sampling \cite{kluck19, asfault19, nsga2sm16, fusionfuzz21, avfuzzer20, asf21, tang321, bussler20}. To improve the coverage of the search space, \cite{nsga2dt18} uses t-wise combinatorial testing\cite{twise04} for the discrete fields and \cite{nsga2dt18, fitest18} select the continuous fields for the initial population by maximizing their average pairwise euclidean distance. Similarly, \cite{autofuzz21} uses random sampling while making sure every sample is a certain distance away from the rest in a metric space.

\noindent\textbf{Evaluation} 
The evaluation step takes in a list of scenario vectors and returns their corresponding fitness scores. The fitness function is designed based on the properties to look for. There are two forms of fitness function: single-dimensional and multi-dimensional. 

A \textbf{single-dimensional fitness function} returns either one objective value or a weighted sum of several objective values. Focus on finding critical scenarios, \cite{kluck19, bussler20, tang321, avfuzzer20} minimize min TTC, min distance, or min distance subtracted by the ego car's minimum stopping distance. \cite{asf21} minimizes the maximum distance the ego car can drive without collision. \cite{asfault19} tries to find wrong lane violations, so it maximizes the ego car's deviation from the lane center. \cite{autofuzz21} designs two fitness functions for finding critical scenarios and wrong-lane scenarios, respectively. It uses min distance, ego car's speed at collision, and NPC vehicle's distance to the ego car's camera's view for the former, and ego car's deviation from lane center as well as its distance to closest wrong lane and sidewalk for the later.
While \cite{fusionfuzz21} also considers min distance, it additionally designs a term accounting for how often the fusion component of the system malfunctions.

A \textbf{multi-dimensional fitness function} returns a multi-dimensional vector with each dimension being an objective value. It is usually preferred when the scenarios are expected to achieve several objectives at the same time.
\cite{nsga2sm16} looks for scenarios where a pedestrian is within the view of an FCW but still collided by the ego car at high speed. Its fitness function thus consists of min TTC, ego car's speed at collision, and min distance for a pedestrian to the FCW's warning area. 
Similarly, \cite{nsga2dt18} tries to find scenarios where an AEB recognizes a moving object but fails to brake and collides with an NPC object at high speed. Its fitness function thus consists of min distance, speed of ego car at collision, and confidence of the detection.
In contrast, \cite{fitest18} aims to find integration component induced failures of an ADAS, which is defined by an unsafe action from one upstream component being selected to dominate a safe action from another upstream component. The fitness function consists of branch coverage of the integration component, failure distance of a certain violation, and unsafe overriding distance of the integration component.

\noindent\textbf{Survival}
After the evaluation, a subset of the scenario vectors is selected at each generation to survive and serve as the candidates for parent selection. Typically, the ones with the best fitness scores among the last generation and the current generation are kept\cite{nsga2sm16, nsga2dt18, fitest18, autofuzz21, fusionfuzz21}. In contrast, other less greedy methods only keep the scenario vectors of the current generation\cite{avfuzzer20, asfault19, tang321}. To increase exploration further, \cite{asf21} only keeps those that can generate new route coverage of the ego car.

\noindent\textbf{Parent Selection}
Parent selection is responsible for selecting the parent scenario vectors used to generate the current scenario vectors. It usually balances between exploration and exploitation. Many works\cite{nsga2sm16, nsga2dt18, fitest18, autofuzz21, fusionfuzz21, tang321} use binary tournament selection\cite{tournamentselection}, which creates duplicates of previous scenario vectors, randomly selects twice the number of needed parents, pairs them up, and selects the one from each pair with a higher fitness score. \cite{avfuzzer20} uses roulette selection which selects parents from the existing scenario vectors with probability proportional to their fitness scores. Since \cite{bussler20} uses DE, it has a very different parent selection step. In particular, it samples three scenario vectors, adds the first two's difference to the third, and selects the third one as a parent paired with the scenario vector from the last generation. Unlike the others, \cite{asf21} does not have parent selection nor crossover. Instead, it keeps a priority queue consisting of scenario vectors to mutate from directly.

\noindent\textbf{Crossover}
A crossover operation randomly mixes two scenario vectors to interpolate two existing scenario vectors. The underlying motivation is that the interpolation of two fit scenario vectors should also be fit. Generic crossover method like simulated binary crossover (SBX)\cite{SBX} is widely used\cite{nsga2sm16, nsga2dt18, fitest18, autofuzz21, fusionfuzz21}. \cite{tang321} uses two-point crossover. Other works adapt the crossover to apply at the level of certain parameters groups. For example, \cite{avfuzzer20} exchanges parameters associated with two NPC vehicles belonging to two scenario vectors. \cite{asfault19} recombines or redistributes road segments of two road networks. 

\noindent\textbf{Mutation}
Mutation is applied to increase diversity. Polynomial mutation\cite{polymutation} is widely used\cite{nsga2sm16, nsga2dt18, fitest18, autofuzz21, fusionfuzz21}. \cite{tang321} uses gaussian mutation. Several other works design more specialized mutations. For example, \cite{avfuzzer20} resamples the value of a randomly chosen field. \cite{asfault19} replaces road segments with new ones. \cite{asf21} designs four types of special mutations for mutating the location of an NPC vehicle.

\noindent\textbf{Additional Steps}
Additional steps have been sometimes used in combination with a basic EA to improve performance further.
Usually, the additional steps are used to filter further or select the scenario vectors generated from the crossover and mutation steps.
For example, \cite{asfault19} uses the Jaccard Index of road segments\cite{Hamers1989SimilarityMI} to filter out similar scenario vectors.
\cite{nsga2dt18} trains a decision tree (DT) classifier to predict critical scenarios every several generations. It filters out those scenario vectors classified into the non-critical leaves, which have less critical scenarios than the non-critical ones.
\cite{nsga2sm16} trains three Neural Networks (NNs) to predict three objective values, respectively. It filters out those scenarios with worse predicted objectives than the previous best scenario vector's objectives. 
\cite{autofuzz21} generates more scenario vectors in the crossover and mutation steps with scenario vectors similar to the already found critical scenarios filtered out and chooses the top ones based on the prediction confidence of an actively trained binary NN classifier predicting critical scenarios. Unlike the previous works, it has an extra mutation step where the top scenario vectors are applied constrained gradient-guided mutations to be made more critical.
Unlike other works, \cite{avfuzzer20} uses two GAs: a global one and a local one. The global one is a regular GA used to promote diversity. The local one aims to promote finding more critical scenarios by using the most critical scenario vectors found so far as the initial population. It also has a restart process that is triggered during the global GA when the fitness score does not improve. This process samples a new initial population and makes sure it is far away from the existing scenario vectors to promote diversity.

\subsubsection{Simulated Annealing}
\label{sec:sa}
\begin{figure}[ht]
\centering
\includegraphics[width=0.6\linewidth]{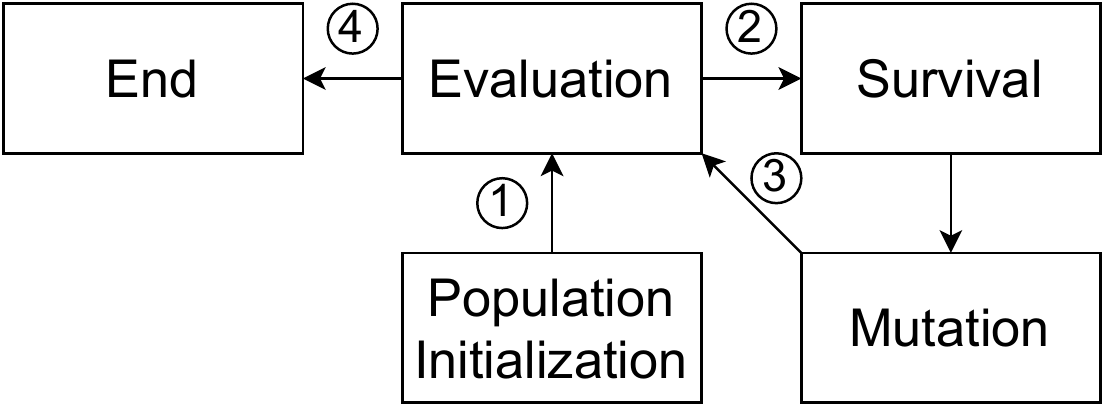}
\caption{The general workflow of a typical SA.}
\label{fig:sa}
\end{figure}

Simulated Annealing is very similar to EA. As shown in \Cref{fig:sa}, the key difference is that SA does not have parent selection and crossover. Besides, it decides whether to keep the original scenario vector at the survival step based on an acceptance function.

\noindent\textbf{Initialization}
\cite{kluck219} uses random sampling. Both \cite{tuncali18} and \cite{simatav20} use t-wise combinatorial testing to generate the initial test cases. The highest ranked scenarios (based on their respective objective functions) are selected.

\noindent\textbf{Evaluation}
The process is similar to GA, except the fitness score is called objective value. \cite{kluck219} focuses on finding critical scenarios and uses min TTC as the only objective. To find critical scenarios which are likely to be the ego car's responsibility, \cite{tuncali18} minimizes min distance, maximizes ego car's speed, and minimizes other objects' distance to the camera view in the objective function. \cite{simatav20} adopts objective function corresponding to its goal of finding requirement violating scenarios (see \Cref{sec:other_property}). 

\noindent\textbf{Survival}
The replacement of each scenario vector by its mutated one is decided based on an acceptance function:
\[ 
  \mathbf{P}(r',r,t) = 
  \begin{cases} 
  1, & \textrm{if } r<r'   \\ 
  exp(\frac{-(r'-r)}{t}), & \textrm{otherwise}
  \end{cases}
\]
where $r'$ is the current objective value, $r$ is the previous objective value, and $t$ is a temperature value which usually linearly decreases as time goes by. The acceptance function essentially says that if the mutated scenario vector has a better objective value, it will replace the original scenario vector. Otherwise, the replacement is based on a probability decreasing over time.

\noindent\textbf{Mutation}
\cite{tuncali18, kluck219, simatav20} mutate a scenario vector by sampling the perturbation from a zero-mean normal distribution.
\cite{kluck219} further shrinks the sphere (from which new scenario vectors are allowed to be sampled from) around the current scenario vector over time as suggested in \cite{locatelli02} to increase convergence speed.

\subsubsection{Coverage and Local Search}
CLS first generates some initial scenario vectors satisfying a certain coverage criterion and then conducts a local search based on the results.  

\cite{paracosm21} tries to cover the search space and finds critical scenarios at the same time. To promote more coverage, it applies t-wise combinatorial testing for the discrete fields and Halton sampling\cite{rote96} for the continuous fields. To promote finding critical scenarios, the algorithm generates scenarios by applying random mutation for those with the smallest min distance so far.

\cite{tang221} aims to find more critical scenarios. It first categorizes routes from a map based on a junction lane pair metric and chooses the representatives from each category as the routes to run. The algorithm then tries to find more critical scenarios by applying a bisection search on parameters associated with NPC vehicles for each representative route with the min distance as the objective function. 

\subsubsection{Autoregressive Model}
\cite{ding20} aims to find more critical scenarios in a cyclist crossing street scenario. The algorithm represents the traffic scenarios with a series of autoregressive building blocks\cite{autoregressive} based on expert knowledge and generates diverse scenarios by sampling from the joint distribution of these blocks. The optimization is formulated in the RL terminologies. In particular, the action space is the scenario vector search space $\mathbf{D}$, and the state space consists of the ego car's route and target speed. The reward is a weighted sum of min distance, if a collision happens, and the distance between the cyclist's initial location to the ego car's projected route. The first two terms promote collisions from happening,  and the last term prevents the cyclist's initial location from being on the ego car's trajectory. Unlike a regular DRL, each state is sampled rather than transited from the last state. Each field of the scenario vector is modeled by a normal distribution, and its conditional probability inference is modeled by a NN taking in state and relevant action variables (based on the expert designed autoregressive model). 

\subsubsection{Bayesian Optimization}
\cite{abeysirigoonawardena19} tries to find a critical scenario using the shortest time in an unprotected right turn scenario. It uses Bayesian Optimization (BO)\cite{boreview} to optimize the policy of an NPC vehicle and a pedestrian. At each time step, the policy decides the location and speed of both NPCs for the next time step. The algorithm uses Gaussian Process (GP) to model the cost function (which predicts the min distance based on the current NPCs' policy) and queries the policy for the next time step by maximizing the expected improvement at the current time step. 

\cite{gangopadhyay19} aims to find more critical scenarios in an occluded pedestrian crossing street scenario. It also applies BO with the acquisition function being the expected improvement in terms of min distance. Its key difference with \cite{abeysirigoonawardena19} is that the GP takes in NPC's initial condition rather than NPC's control policy. Besides, to find more diverse critical scenarios, after finding a critical scenario, it eliminates the corresponding local minimum from the search space and restarts a new search process from the other regions. 

\subsubsection{Other}
\cite{zhu20} conducts an optimization search to find diverse, critical scenarios for an ACC in a car following scenario. The algorithm consists of an exploration stage and an exploitation stage. 
At each round, if the current scenario's min TTC is smaller than the previous scenario, the algorithm enters the exploration stage. Otherwise, it enters the exploration stage or the exploitation stage based on a specified probability.
At the exploration stage, it applies the analytic hierarchy method\cite{qin18} to determine the importance value of each feature. Each field's importance value is then used to determine the mutation magnitude for that field.
At the exploitation stage, the algorithm generates a scenario vector by interpolating the most sparse and dangerous search groups (clustered by DBSCAN\cite{dbscan96}). 

\cite{ding21} tries to find more diverse, critical scenarios in a pedestrian crossing street scenario. It first trains a RealNVP\cite{realnvp16} (a flow-based model for likelihood inference) using a dataset and uses it as a prior generative model. 
At each round, the algorithm applies natural evolution strategy \cite{daan08}, a gradient-based sampling procedure, to minimize min distance as well as the current model's probability to promote finding critical and diverse scenarios. The algorithm then uses all the data so far to update the generative model with weighted likelihood maximization\cite{Wang01} having the weight being a weighted sum of min distance and the scenario's probability in the prior distribution, which essentially weights more on frequently happening, critical scenarios.

\subsubsection{Discussion}
\label{sec:algorithm_sim_adaptive_discussion}
In general, EA and SA are considered scalable with respect to the dimensionality of the search space and have been shown to be effective from several dimensions to several hundred dimensions search space\cite{autofuzz21}. In contrast, both ARM \cite{ding20} and BO \cite{gangopadhyay19} suffer from scalability issues. In particular, \cite{ding20} relies on expert knowledge to design the graph of the autoregressive model, which becomes infeasible when the number of variables grows up. Standard BO has been known to suffer from scalability, although there are recent efforts to apply BO for high dimensional space \cite{pmlr-v70-rana17a, moriconi2020highdimensional}. It might be worth exploring if the scalable BO variations can be applied to \ads testing and outperform EA and SA based algorithms.

\cite{kluck219} compares a standard EA with a standard SA, and shows the superiority of EA. However, it is not clear if this holds in general across scenarios and algorithm variations.
Besides, both EA and SA have many design choices for each procedure, as shown in \Cref{sec:ea} and \Cref{sec:sa}, and it is not clear which design is the best under which setting. A systematic comparison and ablation study is needed. 


\subsection{Step Based Adaptive}
A step-based adaptive algorithm updates an agent controlling certain objects in the simulator (e.g.acceleration of an NPC vehicle) at each time step. We have observed that DRL is the only algorithm in this category.

\subsubsection{Deep Reinforcement Learning}
DRL has been widely used to learn a policy for an NPC vehicle or several NPC vehicles to influence the ego car to be more likely to misbehave over time. Existing works using DRL differ in terms of architecture, state space, action space, and reward.

\begin{figure}[ht]
\centering
\includegraphics[width=0.8\linewidth]{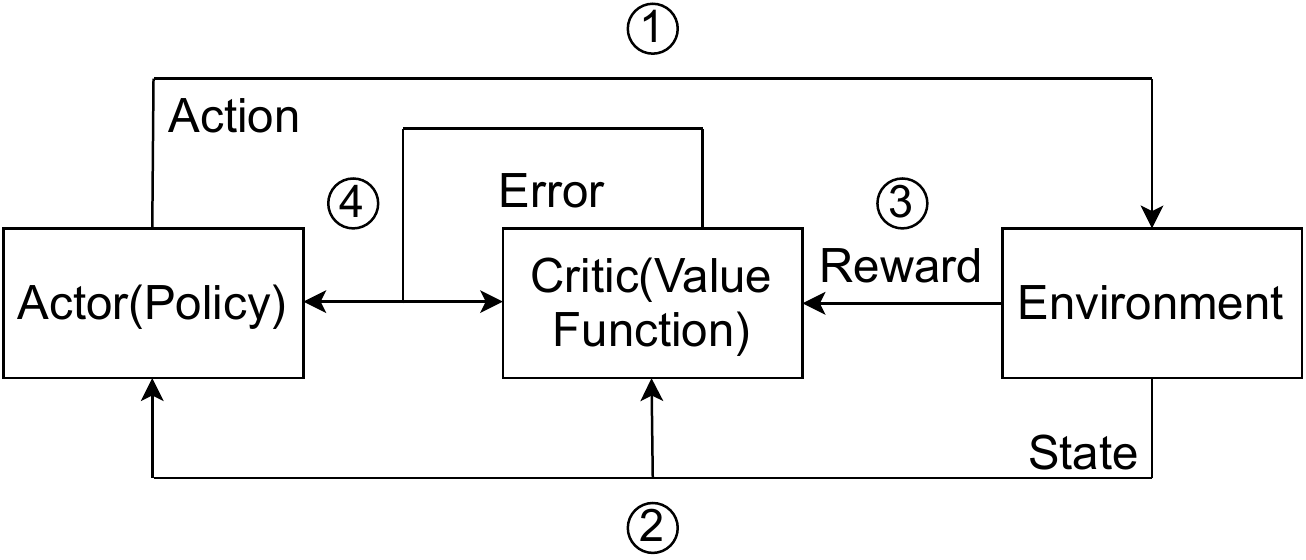}
\caption{The general workflow of a typical RL with actor-critic architecture.}
\label{fig:rl}
\end{figure} 

\noindent\textbf{Architecture} All the works use different variations of an actor-critic architecture (\Cref{fig:rl}). This architecture uses a DNN for policy function (actor) and value function (critic), respectively. At a high level, the workflow of the architecture is: 1. the actor takes an action. 2. the environment updates the state based on the action and the current state. 3. the environment returns a reward to the critic. 4. the critic uses the reward to update its value function and returns the feedback to the actor which also updates its policy. 

Existing works differ in what concrete architecture to use and how to model the actor. In particular, \cite{kuutti20} uses A2C\cite{a2c16} to model a leading NPC vehicle. \cite{failmaker-advrl19} uses Multi-Agent DDPG\cite{multiddpg17} to model each NPC vehicle as a DDPG agent separately. 
\cite{chen21} uses Ensemble DDPG\cite{ddpg16} to model all NPC vehicles with multiple DDPGs to promote diversity. 

\noindent\textbf{State Space}
The state space is usually designed based on the state information of the ego car and NPC vehicles. The state space of \cite{failmaker-advrl19} consists of the locations of all vehicles. \cite{chen21} additionally includes their speed. \cite{kuutti20} additionally includes the acceleration of the ego car while modeling the NPC vehicles' locations and speed in a relative measure with respect to the ego car.

\noindent\textbf{Action Space}
The action space consists of variables through which an agent exerts influence on the simulation environment. It usually consists of the acceleration of NPC vehicles\cite{kuutti20, chen21}. In contrast, the action space of \cite{failmaker-advrl19} is each NPC vehicle's relative distance to a reference trajectory.

\noindent\textbf{Reward}
The reward function is designed to train one or several agents' behavior such that the final scenarios satisfy certain desirable properties. \cite{kuutti20} focuses on finding critical scenarios, so it only uses the min time headway as the reward function. In addition to consider scenarios being critical, \cite{chen21} also considers the ego car's task failure (e.g., failing to finish lane-change). It thus designs the ego car's component reward for NPC vehicles to be
\[ 
\scriptstyle
  -r_{ego} = 
  \begin{cases} 
  -100, & \textrm{if a lane-change is finished}   \\ 
  50, & \textrm{if a collision happened}   \\
  -0.1v_{ego}, & \textrm{otherwise} 
  \end{cases} 
\]
To ensure NPC vehicle follow traffic rules, \cite{chen21} penalizes the traffic-rule violating behaviors of each NPC vehicle. 
\cite{failmaker-advrl19} models each NPC vehicle's behavior separately and assigns more rewards to the one that contributes more to an ego car's collision. To estimate each NPC's contribution, it reruns scenarios where the ego car encounters a collision using different subsets of the NPC vehicles. For each NPC vehicle that has been considered being a contributor, the magnitude of its contribution depends on its min distance to the ego car. 
To promote natural scenarios, \cite{failmaker-advrl19} rewards the NPC vehicle with a smaller distance to a reference trajectory and at a higher speed.


\subsubsection{Discussion}
\label{sec:algorithm_step_adaptive_discussion}

For a DRL method to be efficient for a general application, good designs of architectures, state space, action space as well as the reward function are necessary, if not sufficient. However, in the \ads testing field, it is still unclear what combination of architecture, reward function, state space, and action space is optimal for a given system and scenario since existing works have not provided comprehensive ablation studies nor compared against each other. A potential direction is to conduct an in-depth analysis of these designs and their pros and cons under different settings.


\section{Algorithm for Performance Estimation}
\label{sec:algorithm_performance_estimation}
In this section, we introduce the algorithms used to search parameters in $\textbf{D}$ for performance estimation (\Cref{sec:performance_estimation}). The algorithms are either non-adaptive or simulation-based adaptive. 


\subsection{Non-Adaptive}
\subsubsection{Monte Carlo}
\cite{nitsche18} tries to estimate two AEB systems collision rates and conflict rates in a right-turn at T-junction scenario. It first conducts crash data analysis on an in-depth junction accident dataset and identifies distinct groups of critical scenarios using a clustering method called Partitioning Around Medoids\cite{leonard19}, which is chosen due to its robustness against outliers and ability to cope with categorical data. \cite{nitsche18} then applies Monte Carlo sampling on the search space designed based on a chosen cluster and evaluates the ego car's collision rate and conflict rate.

\subsection{Simulation Based Adaptive}
\subsubsection{Bayesian Optimization}
\cite{zhou18} tries to efficiently find critical class decision boundaries in the search space for an ACC in an NPC vehicle cut-in scenario. It models the probability of a scenario vector inducing a collision using a GP.
At each round, the algorithm samples new scenario vectors with the maximum predictive variance. In other words, the scenario vectors falling into the model's most uncertain region are selected. To speed up the boundary search process, it also limits the new samples which have input predictive probabilities close to the decision boundary (i.e., regions with $50\%$ estimated collision probability).

\subsubsection{Importance Sampling}
Since the naive Monte Carlo method is usually considered prohibitively inefficient in rare-event estimation, \cite{norden20} applies Importance Sampling (IS) to estimate \op's collision rate in a highway scenario. In particular, it applies adaptive multilevel splitting to estimate the optimal importance sampling distribution.
The central idea is to break the estimation of the collision rate into the product of a sequence of conditional probabilities:
\[
\mathbf{P}[f(x)<\epsilon] = \prod_{k=1}^K \mathbf{P}[f(x)<\epsilon_k~|~f(x)<\epsilon_{k-1}]
\]
where $f(x)$ is the min distance and $\epsilon=\epsilon_K...<\epsilon_k<\epsilon_{k-1}<...<\epsilon_0=\infty$ is a sequence of thresholds.
The algorithm gradually decreases $\epsilon_k$ as $k$ increases while keeping the number of scenario vectors having values smaller than $\epsilon_{k-1}$ being equal to $\rho N$ where $\rho\in (0,1)$ is a coefficient, and $N$ is the number of samples at each round. The kept scenario vectors are used to resample $N$ candidate scenario vectors and are then applied random walk (with a specified kernel) for several steps to generate the scenario vectors to run next.

\subsubsection{Other}
\cite{saquib20} tries to estimate critical value for CARLA PID in an NPC vehicle cut-in scenario. The algorithm first generates the initial scenario vectors via stratified sampling, which randomly samples one representative for each region in the search space. After evaluating these scenario vectors, the algorithm samples each new scenario vector with probability proportional to its interpolated critical value (e.g., min TTC). The algorithm finally terminates if the estimated critical values associated with the latest sampled scenario vectors become stable.

\subsection{Discussion}
\label{sec:algorithm_performance_estimation_discussion}
The works on performance estimation are relatively sparse in the context of system-level testing in a high-fidelity simulator. One potential direction is adapting the techniques from the works on component-level testing in a numeric simulation or 2D simulator.
For example, IS and its variations have been widely used in component-level \ads testing \cite{zhao17, huang17, zhao18, huang217, huang317, jesenski20, erwin17, xu21, okelly18, huang417, huang18}. Besides, other techniques like rapid-exploring random trees (RRT)\cite{tuncali19} and variants of sequential Monte Carlo sampling (SMC)\cite{huang517, sinha20} have also been explored at component-level.
It is worth exploring how to apply these methods in the context of system-level testing in a high-fidelity simulator, and seeing how transferable they will be as well as how to potentially adapt them if necessary.

\section{Related Work}
\label{sec:related}
In this section, we compare the current work with some existing surveys related to \ads testing. Note that since the field of \ads testing moves very fast, many works come after these surveys so the overlapping of the reviewed works is relatively limited.

\cite{stellet2015survey} proposes a taxonomy on testing methods for \adas and reviews relevant methods at a high level. \cite{simulationtest2019survey} analyzes the application of simulations for \ads validation based on a 3-circles model\cite{threecircles19} and summarizes relevant works at a high level. In contrast to these two works, we focus on presenting and comparing concrete systems, scenario parameters, testing objectives, and algorithms in depth. \cite{Neurohr20scenariosurvey} provides an overview of each procedure of scenario-based testing and reviews works for each procedure separately. In contrast, we review works that implement the entire testing pipeline. \cite{ScenarioTestSurvey} surveys methods on scenario-based testing for \ads. It compare methods at the category level while we compare methods at the individual work level. \cite{ValidationADSafetySurvey} reviews software verification and validation of autonomous cars. It covers simulation environments and mutation testing, corner cases and adversarial examples, fault injection, software safety cages, techniques for cyber-physical systems, and formal methods. In contrast, we focus on a more specific topic, scenario-based \ads testing in high-fidelity simulation in much more profound depth. \cite{Corso2020ASO} reviews widely used algorithms for black-box safety validation. It focuses on presenting the general algorithms, while we present the algorithms along with other components (e.g., simulators, systems, scenario parameters, and testing objectives) in the context of \ads testing. A concurrent work \cite{zhang2021finding} provides a systematic taxonomy for finding critical scenarios in general. In contrast, we focus on scenario-based testing in high-fidelity simulators.

\section{Challenges and Future Directions}
\label{sec:discussion}


Based on our review of the literature as well as our discussions at the end of each section, we identify the following challenges and future directions regarding different components in \Cref{fig:workflow}.

\noindent\textbf{System and Environment:}
As discussed in \Cref{sec:system_discussion}, a fundamental challenge of using simulation is the gap between the virtual world and the real world. Efforts on quantifying the gaps under different settings and mitigating the gaps are needed. The former can inform a tester what extent of errors (w.r.t. the corresponding testing results in the real world) one can expect from the testing results in a simulation environment. The latter helps to scale up the testing process since testing in the real world is much more expensive than testing in simulation. 

\noindent\textbf{Construct Scenario:} 
As discussed in \Cref{sec:scenario_discussion}, a comparison of different parametrization is highly desirable. In essence, a good search space should optimally trade off the coverage of the functional scenario (through fine-grained parametrization) and tractability (for efficient search). 

\noindent\textbf{Testing Objective:}
As discussed in \Cref{sec:testing_objectives_discussion}, it is important to systematically identify the desirable properties for each setting of ODD and functionality to test as well as understanding their trade-offs in practice.

\noindent\textbf{Algorithm:}
Regarding the search algorithm, many challenges and potential directions have been discussed earlier in \Cref{sec:algorithm_non_adaptive_discussion}, \Cref{sec:algorithm_sim_adaptive_discussion}, \Cref{sec:algorithm_step_adaptive_discussion}, and \Cref{sec:algorithm_performance_estimation_discussion}.
One common observation for all the categories of algorithms is the lack of comparisons and systematic ablation studies. The causes are the limited accessibility of the available \adss or \adass as well as the simulators, and the lack of consensus regarding the algorithm's evaluation metric. A benchmark is needed for researchers and practitioners to compare the existing methods and easily evaluating the performance of newly developed algorithms.

Besides the aforementioned directions, another observation is that the most majority of works reviewed treat the system under test as a black-box. This allows the proposed method to be system independent. On the flip side, it also limits their use of the system internal information during the simulation, which might greatly aid the performance improvement of the algorithm as well as the analysis of the root causes. One potential direction is to apply some white-box or grey-box testing techniques for the same problems.

\section{Conclusion}
\label{sec:conclusion}
In this work, we study a generic formulation of scenario-based testing in high-fidelity simulators and conduct a literature review on relevant works. We further discuss the open challenges and promising research directions. We hope this work can provide a nice introduction to the state-of-the-art on this topic and inspire future work.

\ifCLASSOPTIONcaptionsoff
  \newpage
\fi



%

\bibliographystyle{IEEEtran}

\bibliography{bib_main}


%








\end{document}